\documentclass[./subfig]{./aa}

\input epsf

\newcommand{\bb}{\bibitem[]{bla}}
\newcommand{\zm}{ \relax \ifmmode {\rm M_{\odot}} \else {M$_{\odot}$}\fi}

\newcommand{\ang}{$\rm \AA$}

\newcommand{\degree}{$^{\rm o}$}

\newcommand{\mic}{$\mu$m}
\newcommand{\ea}{{et al.}}

\newcommand{\ha}{H$\alpha$}

\def\lesssim{\mathrel{\hbox{\rlap{\hbox{\lower4pt\hbox{$\sim$}}}\hbox{$<$}}}}
\def\gtrsim{\mathrel{\hbox{\rlap{\hbox{\lower4pt\hbox{$\sim$}}}\hbox{$>$}}}}

\def\ion#1#2{#1$\;${\small\rm\@Roman{#2}}\relax}
 
\newbox\grsign \setbox\grsign=\hbox{$>$} \newdimen\grdimen 
\grdimen=\ht\grsign
\newbox\simlessbox \newbox\simgreatbox
\setbox\simgreatbox=\hbox{\raise.5ex\hbox{$>$}\llap
     {\lower.5ex\hbox{$\sim$}}}\ht1=\grdimen\dp1=0pt
\setbox\simlessbox=\hbox{\raise.5ex\hbox{$<$}\llap
     {\lower.5ex\hbox{$\sim$}}}\ht2=\grdimen\dp2=0pt

\makeatletter 
\renewcommand\@biblabel[1]{}     % Arabic numbers, 

\makeatother

\begin{document}

\title{EXPORT:  Optical photometry and polarimetry of Vega-type and pre-main sequence stars 
\thanks{Table A1 will only be electronically available at CDS via anonymous ftp to \tt cdsarc.u-strasbg.fr}
}

\author{Ren\'e D. Oudmaijer\inst{1}, J. Palacios\inst{11}, C. Eiroa\inst{11},
J. K. Davies\inst{8},  
D. de Winter\inst{4}, R. Ferlet\inst{5}, F. Garz\'on\inst{2}, 
C. A. Grady\inst{6}, 
A. Cameron\inst{7},  H. J. Deeg\inst{3}, 
A. W. Harris\inst{7,9}, K. Horne\inst{7}, B. Mer\'\i n\inst{10}, 
L. F. Miranda\inst{3}, B. Montesinos\inst{3,10}, A. Mora\inst{11},
A. Penny\inst{12}, 
A. Quirrenbach\inst{13}, H. Rauer\inst{9}, J. Schneider\inst{14}, 
E. Solano\inst{10}, Y. Tsapras\inst{7}, P.R. Wesselius\inst{15}
 }
\offprints{R.D. Oudmaijer}

\institute{The Department of Physics and Astronomy, E C Stoner Building,
Leeds, LS2 9JT,  UK
\and
Instituto de Astrof\'\i sica de Canarias, 38200 La Laguna, Tenerife,
Spain
\and
Instituto de Astrof\'\i sica de Andaluc\'\i a,
Apartado de Correos 3004, 18080 Granada, Spain
\and
TNO/TPD-Space Instrumentation, Stieltjesweg 1, PO Box 155,
2600 AD Delft, The Netherlands
\and
CNRS, Institute d'Astrophysique de Paris, 98bis Bd. Arago,
75014 Paris, France
\and
NOAO/STIS, Goddard Space Flight Center, Code 681, NASA/GSFC,
Greenbelt, MD 20771, USA
\and
Physics \& Astronomy, University of St. Andrews,
North Haugh, St.Andrews KY16 9SS, Scotland, UK
\and
Joint Astronomy Centre, 660 N. A'ohoku Place, Hilo, Hawaii 96720,
USA
\and
DLR Institute of Space Sensor Technology and 
Planetary Exploration, Rutherfordstrasse 2,
12489 Berlin, Germany
\and
LAEFF, VILSPA, Apartado de Correos 50727, 28080 Madrid, Spain
\and
Dpto. F\'\i sica Te\'orica, C-XI, Facultad de Ciencias, 
Universidad Aut\'onoma de Madrid, Cantoblanco, 28049 Madrid, Spain 
\and
Rutherford Appleton Laboratory, Didcot, Oxfordshire OX11 0QX, UK
\and
Department of Physics, Center for Astrophysics and Space Sciences,
University of California San Diego, Mail Code 0424, La Jolla, CA 92093-0424,
USA
\and
Observatoire de Paris, 92195 Meudon, France
\and
SRON, Universiteitscomplex ``Zernike", Landleven 12,
P.O. Box 800, 9700 AV Groningen, The Netherlands
}

\date{received,  accepted}

\abstract{ This paper presents optical {\it UBVRI} broadband
photo-polarimetry of the EXPORT sample obtained at the 2.5m Nordic
Optical Telescope.  The database consists of multi-epoch
photo-polarimetry of 68 pre-main-sequence and main-sequence
stars.  An investigation of the polarization variability indicates
that 22 objects are variable at the 3$\sigma$ level in our data. All
these objects are pre-main sequence stars, consisting of both T Tauri
and Herbig Ae/Be objects while the main sequence, Vega type and post-T
Tauri type objects are not variable. The polarization properties of
the variable sources are mostly indicative of the UXOR-type behaviour;
the objects show highest polarization when the brightness is at
minimum. We add nine new objects to the class of UXOR variables (BH
Cep, VX Cas, DK Tau, HK Ori, LkH$\alpha$ 234, KK Oph and RY Ori). The
main reason for their discovery is the fact that our data-set is the
largest in its kind, indicating that many more young UXOR-type
pre-main sequence stars remain to be discovered.  The set of Vega-like
systems has been investigated for the presence of intrinsic
polarization. As they lack variability, this was done using indirect
methods, and apart from the known case of BD +31\degree643, the following
stars were found to be strong candidates to exhibit polarization due
to the presence of circumstellar disks: 51 Oph, BD +31\degree643C, HD
58647 and HD 233517.
 } 

\authorrunning{R.D. Oudmaijer et al. EXPORT}
\titlerunning{Photo-polarimetry of young stars }

\maketitle

\section{Introduction}

One of the main projects within the EXPORT project (EXoPlanetary
Observational Research Team, Eiroa et al. 2000)
concerned the study of the circumstellar gas and dust around
pre-main-sequence (PMS) and main-sequence (MS) objects thought to be
accompanied by planets, or by planets in formation.  The aim of this
effort was to study stars at different early evolutionary stages by
gathering more observational data on the existing samples in order to
provide better defined observational clues regarding proto-planetary
disk evolution.  The data concerned are near-simultaneous
spectroscopy, photometry and polarimetry. Such data are complementary,
as, roughly speaking, the former data probe the gas closest to the
stars, while the photo-polarimetry most likely probes gas and dust
further out from the stars.  Although studies of individual objects
have been carried out in the recent past, a large sample of objects
has not been studied simultaneously in a systematic and statistical
manner. This is mainly due to the large amounts of telescope time
needed -- made possible only now by the allocation of the 1998 La
Palma International Time to the EXPORT project.
Near-infrared photometry of our sample has been published in Eiroa et
al (2001), while the first results on the spectroscopy are presented
in Mora et al (2001).  

This paper concerns the presentation and preliminary analysis of the
optical photo-polarimetry obtained in the course of the project.  The
purpose of these data is twofold, in the first place, the data are
used to `flux' the spectroscopic data so that not only changes in the
emission line equivalent widths can be traced, but also the energy
output. Secondly, combined polarimetry and photometry teaches us a
great deal about the geometry of the circumstellar material. Stellar
light that is scattered off circumstellar dust grains will result in
the light being polarized. In the case of spherically symmetric
shells, or face-on circular disks, no net polarization will be
observed as all polarization vectors cancel each other out. On the
other hand, in the cases of a deviation from circular symmetry on the
sky, a net polarization will be observed.

It is not trivial to assess whether a polarized object indeed has a
disk-like dust configuration, as interstellar polarization can often
contribute significantly to the observed polarization.
Variability of the polarization can settle this question quite
straightforwardly, as any variations in polarization must be due to
circumstellar effects, because interstellar dust clouds are not expected to
vary within the epochs under consideration.

A sub-sample of both low and intermediate mass young objects displays
a special type of photo-polarimetric variability, commonly referred to
as the `UX Ori' phenomenon (Grinin et al 1994).  From long term
photo-polarimetric monitoring of a small number of young stars, Grinin
et al (1994) identified a group of objects that are
photo-polarimetrically variable. This group of stars shows increased
polarization when the optical light of the stars is faint. Crucially,
the objects also are redder at fainter magnitudes, while in extreme
visual minima there is a colour reversal, the observed colours become
bluer again.  Named after their proto-type, UX Orionis, these stars
are commonly referred to as UXORs. UX Ori itself is classified as a
Herbig Ae/Be star, and indeed many UXORs are on the more massive end
of the young stellar mass spectrum. Yet, this behaviour is not
confined to Herbig stars, as for example the T Tauri object BM And
does show the same behaviour (Grinin et al 1995).

The main explanation of this phenomenon concerns the existence of dust
clumps located in a disk-like configuration rotating around the star.
When the dust clumps are not in our line of sight, the star will be
observed at maximum light, with only a slight contribution of
scattered radiation from the dust.  If the dust is intersecting the
line of sight, light from the star will be absorbed, and the relative
contribution of the scattered light increases, increasing the observed
polarization. Moreover, the fact that the reddening of the star
coincides with the faintening, leaves little doubt that
dust-absorption indeed plays the main role in the process. In cases of
extremely deep minima, the light from the star is blocked almost
entirely, resulting in a `blueing' of the energy distribution, as now
mostly scattered light dominates the observed light.  Depending on the
distribution of dust-clouds, the light can be more or less absorbed
during a period of photo-polarimetric monitoring. Added complications
to this simplified scenario are aligned grains and dust production and
destruction (see Grinin 1994).  For more detailed discussions on the
origins of UXOR photo-polarimetric behaviour, we refer the reader to
Grinin (1994) and Grinin et al (1994), but see Herbst \& Shevchenko
(1999) who discuss some short-comings of the scenario.  In the
following, we will refer to stars exhibiting the UX Ori phenomenon
when an increased polarization is observed simultaneously with
fainter, and possibly redder, photometric results. We appear not to
have observed objects in very deep minima, hence the `blueing effect'
is not present in our data.

The aims of this paper are threefold. It is meant as a database for
the community, and as such the data-acquisition, reduction and results
are presented. In addition, the first analysis of the data is
presented, aimed firstly at the detection of variability of the
sources, and secondly at identifying Vega-type objects that may exhibit
intrinsic polarization due to dusty circumstellar circumstellar disks.

This paper is organized as follows: Sect.~2 presents the observations
and data reduction, Sect. 3 discusses the variability of the
sources and Sect. 4 presents the analysis of the Vega-type objects,
while we end with a discussion in Sect.~5.

\section{Source selection, observations and data-reduction}

\subsection{Source Selection}

In order to learn more on the statistical behaviour of planetary
systems, and especially their formation, the observed sample covers
stars in various stages of their evolution.

Low mass young stellar objects such as the Classical T Tauri stars
(CTTs) and Early type T Tauri stars (ETT) were selected from Herbst et
al (1994). A sample of their intermediate mass counterparts, the
Herbig Ae/Be stars, was taken from the Th\'e et al (1994)
catalogue. The observed sample covers almost all such stars observable
in the Northern hemisphere.  Objects further in their evolution, which
are now on, or close to, the Main Sequence, were also included.  This
sub-sample includes Vega-type stars, MS stars that are found to exhibit
an infrared excess identified with circumstellar, possibly
proto-planetary, disks. The spectral  types cover the low
to intermediate mass range. These objects were mostly taken from
Sylvester et al (1996).  Finally, some A-type shell stars found
to be accreting gas, a spectral characteristic of the Vega-type object
$\beta$ Pic, were taken from Grady et al (1996). The list of targets
and some results are presented in Table~\ref{log}.

\subsection{Observations}

The observations employed the Turpol {\it UBVRI}
polarimeter/photometer (Korhonen et al. 1984) mounted on the the 2.5m
Nordic Optical Telescope (NOT), La Palma, Spain.  The data are
obtained simultaneously in the {\it UBVRI} bands (with equivalent
wavelengths of 0.36, 0.44, 0.53, 0.69 and 0.83 \mic ) via several
dichroic beamsplitters. A half-wave plate rotated in steps of
22.5\degree \ enables linear polarization measurements to be made.
One complete polarization measurement consists of eight integrations.
The sky background polarization is eliminated by the use of the
calcite block, so only sky intensity measurements are recorded, mostly
before and immediately after a target observation.

The data were collected on photo-multipliers. Neutral density filters
(1\% and 10\% throughput respectively) were used to block the light
from very bright sources. The use of these neutral density filters
resulted in roughly the same number of photons per unit time being
recorded from bright sources and faint sources.  Hence, for the same
exposure times, similar observational error bars on the resulting
polarization and photometry are obtained for the targets.  The objects
have been observed with exposure times of 10 sec (per integration)
with 4 cycles giving total exposure times of 320 sec, with the
exception of the photometric standards for which 1 cycle only was
taken.

The observations were obtained during the nights of 14-17 May 1998,
28-31 July 1998, 23-27 October 1998 and 29-31 January 1999.  All
nights were of good enough quality to yield useful polarization data,
while about half of the nights were photometric.  Apart from July, in
all runs a 10 arcsec diaphragm was used, in July the 7.5 arcsec
diaphragm was employed.  Calibration observations were taken
throughout the nights, these include polarized standard stars,
zero-polarization stars and photometric standards with different
colours, observed at different air-masses.

\begin{table}
\caption{
The observed objects. Columns 2 and 3 list the number of polarimetric 
and photometric observations respectively. Column 4 gives the weighted mean of the {\it V} band polarization, as an indication of the average level of polarization observed, and finally, it is noted whether the polarization was found to be variable or not (see text).
\label{log}  }
{\scriptsize
\begin{tabular}{lcccccl}
\hline
\hline
Name     & $N_{\rm pol}$ & $N_{\rm phot}$ & $P_V$ (\%)  & Variable? &\\
\hline
\hline
{\bf Herbig Ae/Be:}\\ 
\hline
    AS 442 & 3& 0 & 2.93  $\pm$ 0.04&          & \\
 BD+40\degree 4124 & 4& 3 & 1.17  $\pm$ 0.03&          & \\
  HD 31648 & 6& 5 & 0.40  $\pm$ 0.02&          & \\
  HD 34282 & 3& 3 & 0.12  $\pm$ 0.02&          & \\
 HD 144432 & 7& 3 & 0.25  $\pm$ 0.02&          & \\
 HD 150193 & 3& 2 & 4.64  $\pm$ 0.02&          & \\
 HD 179218 & 1& 1 & 0.64  $\pm$ 0.07&          & \\
 HD 190073 & 3& 1 & 0.41  $\pm$ 0.03&          & \\
  LkH$\alpha$ 234 & 8& 5 & 0.44  $\pm$ 0.07& $\surd$  & \\
   MWC 297 & 1& 1 & 1.15  $\pm$ 0.17&          & \\
    VX Cas & 9& 7 & 0.69  $\pm$ 0.03& $\surd$  & \\
    SV Cep &10& 6 & 1.03  $\pm$ 0.02& $\surd$  & \\
 V1686 Cyg &11& 7 & 3.14  $\pm$ 0.07& $\surd$  & \\
    VY Mon & 3& 2 & 9.82  $\pm$ 0.23& $\surd$  & \\
    51 Oph & 6& 3 & 0.47  $\pm$ 0.02&          & \\
    KK Oph & 9& 3 & 5.26  $\pm$ 0.05& $\surd$  & \\
    BF Ori & 6& 4 & 0.46  $\pm$ 0.01& $\surd$  & \\
     T Ori & 6& 4 & 0.29  $\pm$ 0.02&          & \\
    UX Ori & 9& 6 & 1.15  $\pm$ 0.01& $\surd$  & \\
  V346 Ori & 5& 3 & 0.22  $\pm$ 0.02&          & \\
  V350 Ori & 5& 3 & 2.57  $\pm$ 0.05& $\surd$  & \\
   XY Per  & 6& 5 & 1.58  $\pm$ 0.01& $\surd$  & \\
    VV Ser &11& 7 & 1.72  $\pm$ 0.04&          & \\
    RR Tau & 7& 5 & 1.38  $\pm$ 0.03& $\surd$  & \\
    WW Vul &11& 7 & 0.27  $\pm$ 0.02& $\surd$  & \\
\hline
{\bf Herbig/ZAMS:}\\ 
\hline
  HD 58647 & 3& 1 & 0.16  $\pm$ 0.02&          & \\
 HD 141569 & 9& 4 & 0.62  $\pm$ 0.02&          & \\
 HD 142666 & 7& 3 & 0.71  $\pm$ 0.04&          & \\
 HD 163296 & 7& 3 & 0.02  $\pm$ 0.01&          & \\
 HD 199143 & 2& 2 & 0.11  $\pm$ 0.04&          & \\
\hline
{\bf T Tauri:}\\ 
\hline
 HD 123160 & 5& 2 & 0.30  $\pm$ 0.04&          & \\
    BM And & 8& 7 & 2.03  $\pm$ 0.06& $\surd$  & \\
    BH Cep &11& 8 & 0.44  $\pm$ 0.03& $\surd$  & \\
    BO Cep & 9& 6 & 0.88  $\pm$ 0.03&          & \\
    CO Ori & 8& 6 & 2.22  $\pm$ 0.03& $\surd$  & \\
    HK Ori & 7& 6 & 1.18  $\pm$ 0.03& $\surd$  & \\
    NV Ori & 7& 5 & 0.58  $\pm$ 0.01& $\surd$  & \\
    RY Ori & 7& 5 & 2.62  $\pm$ 0.05& $\surd$  & \\
    CQ Tau & 7& 5 & 0.20  $\pm$ 0.01& $\surd$  & \\
    CW Tau & 8& 7 & 0.79  $\pm$ 0.12& $\surd$  & \\
    DK Tau & 5& 5 & 1.31  $\pm$ 0.07& $\surd$  & \\
    DR Tau & 4& 4 & 0.46  $\pm$ 0.06&          & \\
    RY Tau & 7& 6 & 2.62  $\pm$ 0.02& $\surd$  & \\
    PX Vul & 9& 6 & 3.89  $\pm$ 0.03&          & \\
\end{tabular}
\begin{tabular}{lcccccl}
\phantom{\bf Herbig/ZAMS:}     & \phantom{$N_{\rm pol}$} & \phantom{$N_{\rm phot}$} & \phantom{$P_V$ (\%)}  & \phantom{Variable?} & \\
\hline
{\bf Vega/PTT:}\\ 
\hline
  BD+31\degree 643 & 8& 6 & 1.26  $\pm$ 0.03&          & \\
 BD+31\degree 643C & 3& 2 & 0.70  $\pm$ 0.03&          & \\
  HD 23362 & 3& 1 & 0.66  $\pm$ 0.04&          & \\
  HD 23680 & 2& 1 & 1.18  $\pm$ 0.06&          & \\
  HD 34700 & 2& 2 & 0.35  $\pm$ 0.06&          & \\
 HD 109085 & 1& 1 & 0.05  $\pm$ 0.04&          & \\
 HD 142764 & 3& 2 & 1.56  $\pm$ 0.03&          & \\
 HD 233517 & 3& 1 & 1.64  $\pm$ 0.03&          & \\
     HR 26 & 1& 1 & 0.09  $\pm$ 0.06&          & \\
    HR 419 & 3& 1 & 0.19  $\pm$ 0.06&          & \\
   HR 1369 & 1& 1 & 0.03  $\pm$ 0.05&          & \\
  HR 1847A & 3& 3 & 0.63  $\pm$ 0.04&          & \\
  HR 1847B & 3& 3 & 0.63  $\pm$ 0.02&          & \\
   HR 2174 & 3& 1 & 0.07  $\pm$ 0.03&          & \\
  HR 2174B & 1& 0 & 0.07  $\pm$ 0.04&          & \\
  HR 4757B & 1& 0 & 0.11  $\pm$ 0.09&          & \\
  HR 5422A & 4& 2 & 0.08  $\pm$ 0.02&          & \\
  HR 5422B & 4& 2 & 0.06  $\pm$ 0.04&          & \\
   HR 9043 & 3& 3 & 0.04  $\pm$ 0.04&          & \\
$\lambda$ Boo & 4& 1 & 0.03  $\pm$ 0.01&          & \\
    49 Cet & 2& 0 & 0.15  $\pm$ 0.06&          & \\
\hline
{\bf A-shell:}\\ 
\hline
    24 CVn & 3& 1 & 0.02  $\pm$ 0.02&          & \\
    17 Sex & 3& 1 & 0.07  $\pm$ 0.03&          & \\
     HR 10 & 3& 3 & 0.07  $\pm$ 0.04&          & \\
\hline
\hline
\end{tabular}
}
\end{table}

\subsection{Data reduction}

The polarization data were reduced using software designed for the
Turpol, both RDO and JP (using a similar programme) reduced the data
independently, and the results were similar.  1$\sigma$ error bars are
calculated from least-square fits to the data in the eight different
orientations of the retarder, or the photon noise, whichever is the
largest. The resulting accuracy in the Q and U Stokes vectors is of
order 0.05-0.1\% for the sources in most bands. The error in the
polarization is the same as the error in the Stokes vectors. Typical
errors on the polarization angle were of order 1-3\degree \ and
calculated from 0.5 $\times$atan($\sigma_P/P$), which, to first
approximation equals the commonly adopted 28.65\degree
$\times\sigma_P/P$
\footnote{These approximations do not strictly hold for very low
polarization values (see Serkowski 1962, p. 305), but only for $P>>$
0\%, or, roughly, $P \geq$ 6 $\sigma_P$. For $P \sim$ 0\%, the errors
on the polarization are actually smaller than the error bars on Q and
U by about 35\%, while the error bars on the polarization angle reach
a maximum of 52\degree .  These values are related to the complex form
of error-propagation -- see Serkowski 1962, p.304 and further }. In
one instance, during the first two observations of 29 January 1999, a
problem with the field rotator occurred, resulting in an error in the
position angle that is hard to quantify. For these two objects (BD +31
643 and HD 23362), we set the error in PA to 10\degree . It should be
noted here that the computed error-bars reflect the statistical,
internal, error-bars. The external errors are expected to be slightly
larger, of order 0.1\%, as polarization standard stars often have
slightly different published values, possibly due to different
filter-bands and systems used. We also find deviations at the
1-2\degree \ level when intercomparing polarized standards. In those
cases we took the weighted mean of the correction to be applied. This
adds an additional uncertainty to the derived position angles of
1-2\degree , which does not affect the results in the vast majority of
cases.

The instrumental polarization as determined from
observations of zero-polarization stars, was mostly below 0.05\% and
thus not of relevance to the data reported here. 
The system was extremely stable during the four runs, statistical tests,
outlined below, showed that highly polarized standard stars were
recorded with the same polarization throughout, which was consistent with
values from the literature. The zero-polarized stars remained constant
within the error bars, indicating that the instrumental polarization
was constant at low values.

We note that measured polarizations, by virtue of the way they are
calculated ($ P = \sqrt{Q^2 + U^2} $), are always positive, and hence
at the low polarization end, the data are inevitably biased towards
too high polarization values. The usual manner of dealing with such
data is correcting in the manner $P_{\rm true} = \sqrt{P_{\rm obs}^2 - 0.5\pi
\sigma_{P}^2}$ (cf. Serkowski 1962).  Here, we will {\it not} correct
for this bias and {\it not} subtract the instrumental polarization
from the resulting data (mostly due to pragmatic reasons as the
objects often have larger polarizations), but it should be kept in
mind when investigating low polarization targets.

Nine out of the 16 nights were photometric, these were 15 and 16 May,
29 July and 23-27 October 1998, and 30 January 1999.  The photometry
was reduced using standard procedures, and taking into account
atmospheric extinction. It was found that the neutral density (ND)
filters were not entirely grey, and additional corrections had to be
performed on the final values. The problem stemmed from the fact that
the observed Landolt standard stars were mostly faint ({\it V} $>$ 9),
and no ND filter was needed, while some bright target stars needed ND
filters in order to avoid burning the photo-multipliers. The
corrections were determined by comparing existing photometry, taken
from {\sc SIMBAD}, and the results were as follows: For the 10\% ND
filter no correction was needed in the {\it UBVR} bands, while the
correction amounted to --0.15 mag in the {\it I} band. The 1\% ND
filter had a larger effect; --0.08 mag ({\it V}), --0.18 mag ({\it R})
and --0.45 mag in the {\it I} band, while the {\it U} and {\it B}
bands did not need a correction.  The error-bars on the photometry
were dominated by the solutions to the colour transformation rather
than shot-noise statistics, and are less than 0.1 mag overall.
Conservative limits of 0.10, 0.10 and 0.15 mag are adopted for the
{\it V, R} and {\it I} band data respectively where corrections had to
be applied because of the 1\% ND filter and 0.10 mag for the {\it I}
band in the case of the 10\% ND filter.

In total, 68 targets were observed. We attempted to observe every star
at least twice during each observing run, and during consecutive
observing runs. The resulting data set thus contains
photo-polarimetric monitoring on daily and monthly time-scales with
most of the objects having been observed 5-10 times. The resulting
data set is provided in the Appendix, which is  available
electronically at CDS.

\section{Results}

Although a detailed analysis of the data is beyond the scope of this
paper, we discuss some global properties of the sample below.  The
main property that we discuss first is the presence of polarimetric
variability - which confirms the presence of a flattened structure
around the objects. We then discuss the polarization characteristics
of the Vega-type stars and identify  targets for follow-up
research.

\subsection{Variability of the sources}

In this section we will discuss the variability of the sources.
Determining whether the sources are variable is not necessarily
straightforward as outlined in the previous footnote. The error
propagation from the QU vectors is not trivial, complicating
statistical tests on the resulting polarization and angle.  Hence, we
have to perform a test in QU space.

To this end, we adopt the simple, yet powerful statistical formalism
presented by Oudmaijer \& Bakker (1994) and Henrichs \ea\,(1994).  The
variability can be expressed as

\begin{equation}
\left(\frac{\sigma_{\rm obs}}{\sigma_{\rm av}}\right)_{\lambda} \approx 
\sqrt{\frac{1}{N-1} \sum_{i=1}^{N}\left(\frac{ P_{i}(\lambda) -
P_{\rm av}(\lambda)}{\sigma_{i}(\lambda)}\right)^{2}}
\end{equation}

Where $\sigma_{\rm obs}$ represents the scatter of the data-points around
the mean observed value, $\sigma_{\rm av}$ is the (average) experimental
error, $N$ is the number of data points, $P_{av}(\lambda)$ represents
the (weighted) average Stokes parameters, $P_{i}(\lambda)$ the
individual Stokes parameters, and $\sigma_{i}(\lambda)$ is the
1$\sigma$ error-bar for the data points.

($\sigma_{\rm obs}/\sigma_{\rm av}$) corresponds to the standard
deviation of the variations of the individual data points with respect
to the average divided by the standard deviation of the average.  If
no significant variations are present, the value will be close to one,
significant deviations are directly represented in units of the noise
level, that is to say a peak of three corresponds to a variability at
a 3$\sigma$ level.  This method gives a numerical answer to the
question: is the r.m.s. of the mean of all observations larger than
the individual error-bars?  The method was applied on the individual Q
and U vectors for every photometric band of the observed targets. The
mean value was close to 1, confirming that the error-bars provided by
the reduction software reflect the true errors. A more stringent
confirmation of the strength of this statistical test is that the
polarization and photometric standard stars appeared not to be
variable at a level larger than 2$\sigma$. An exception to this
finding was the polarized standard HD 154445. This star was flagged as
variable at the 3$\sigma$ level in almost all photometric
bands. Inspection of the data revealed that this was due to the star
being very close to the bright limits of the photo-multipliers for its
ND filter, which resulted in more counts than recorded for the average
target. As such this star has very low experimental error bars,
sometimes at the 0.03\% and 0.3\degree \ level, so that systematic
uncertainties in polarization angles can not be ignored anymore (see
the discussion in the previous Section).

We selected all objects that displayed variations at the $>$ 3$\sigma$
level in either the Q or U Stokes vector in at least 1 band. These
objects are indicated in Table~\ref{log} \footnote{Four objects that
satisfied this criterion are not listed in the table. The statistical
results of HD 144432, HD 163296 and V346 Ori were dominated by one
deviant polarization value in only one band in the sequence of
observations, which was probably related to the low polarization of
the sources. These stars are not further discussed here. $\tau$ Boo is
effectively unpolarized in our data, with the listed polarizations of
order the observational error-bars.  }. Virtually all T Tau stars were
found to be variable, while half of their earlier-type counterparts,
the Herbig Ae/Be stars are variable in our data-set. There are
undoubtedly more variable stars in the sample, as for example low
level but periodic variability will not be recognized as such using
the above procedure. A detailed investigation however is beyond the
scope of this paper.
\begin{subfigures}
\begin{figure*}

\mbox{\epsfxsize0.32\textwidth\epsfbox[0 100 600 685]{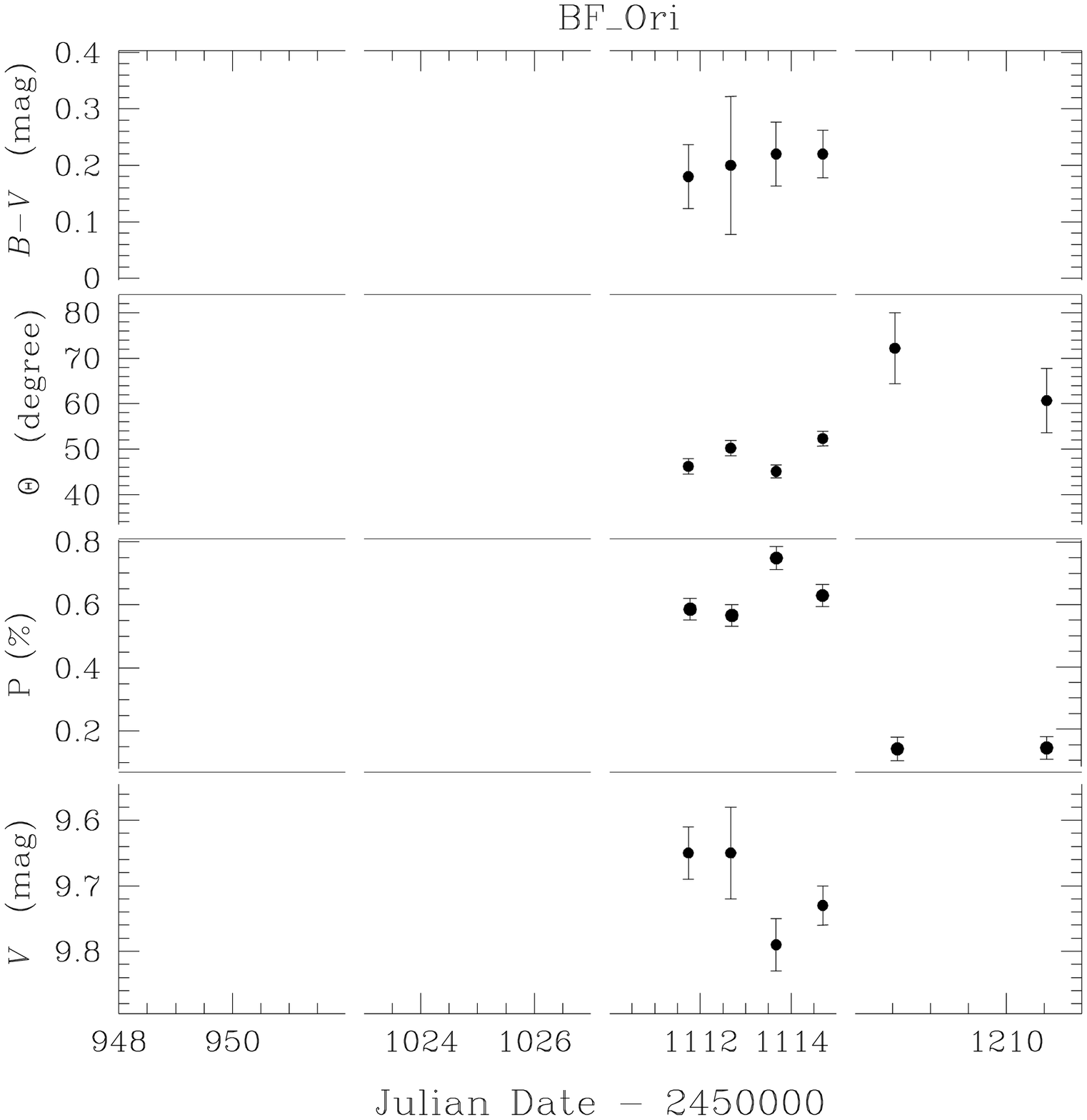}}
\mbox{\epsfxsize0.32\textwidth\epsfbox[0 100 600 685]{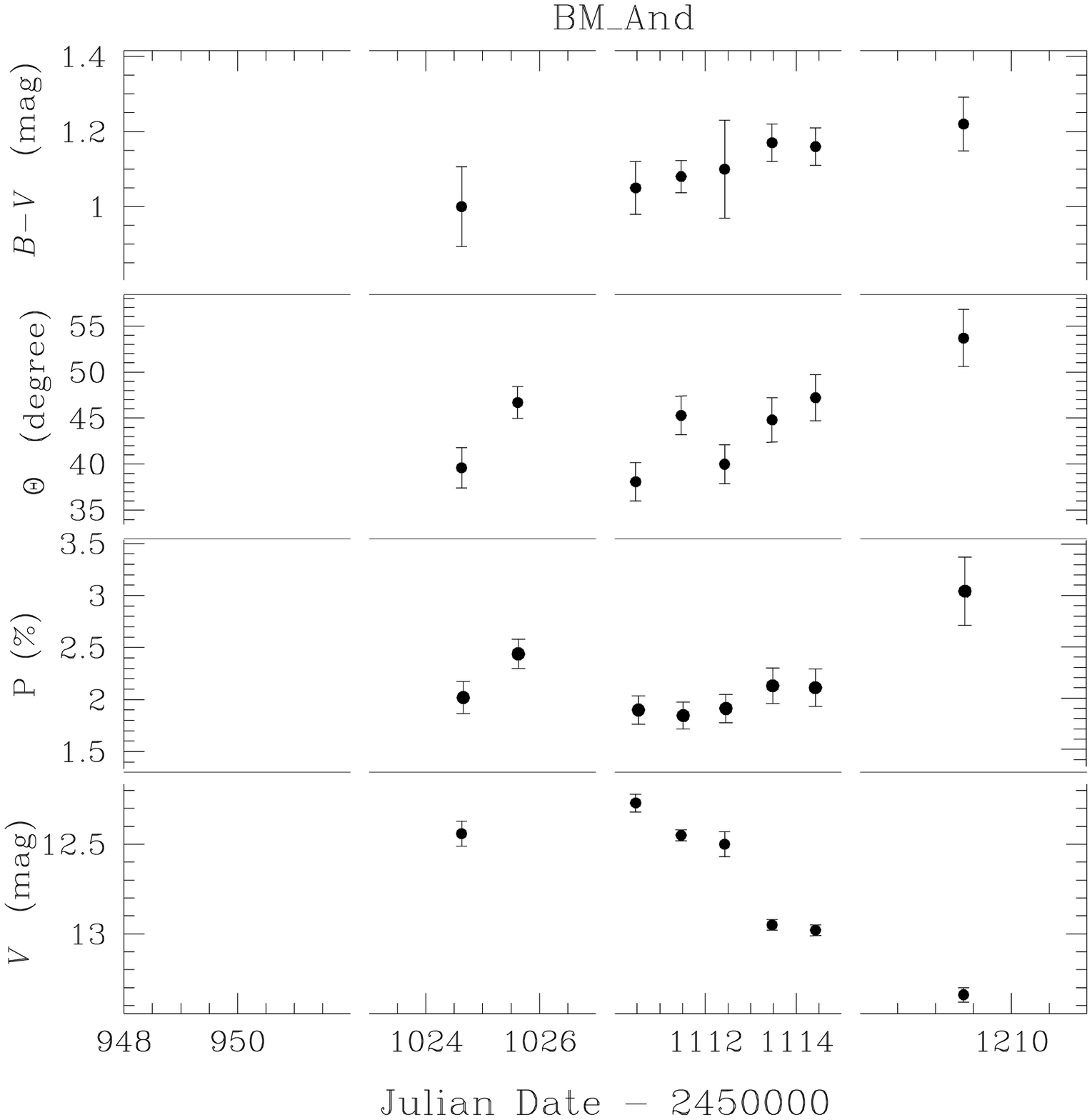}}
\mbox{\epsfxsize0.32\textwidth\epsfbox[0 100 600 685]{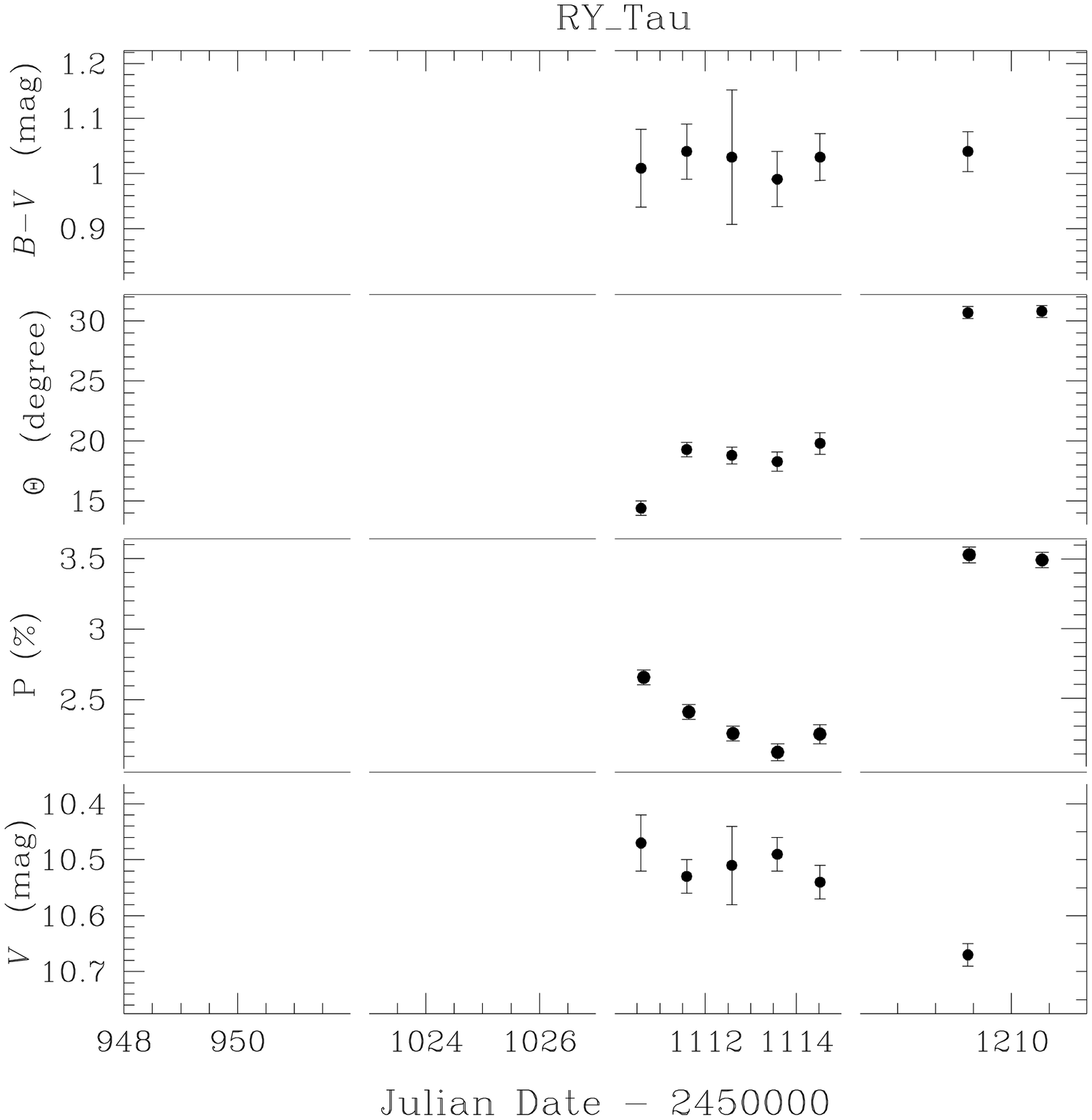}}

\mbox{\epsfxsize0.32\textwidth\epsfbox[0 125 600 685]{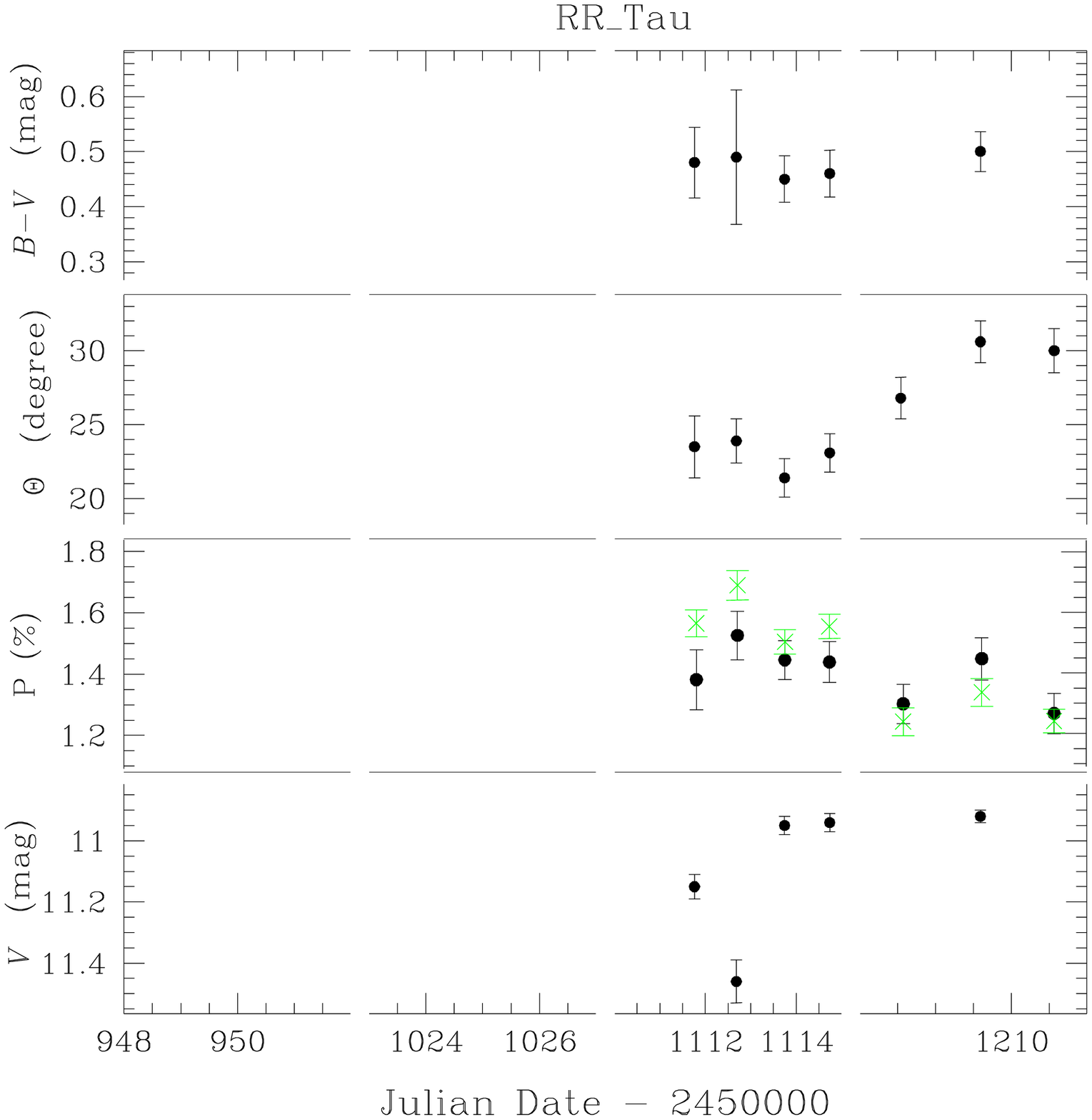}}
\mbox{\epsfxsize0.32\textwidth\epsfbox[0 125 600 685]{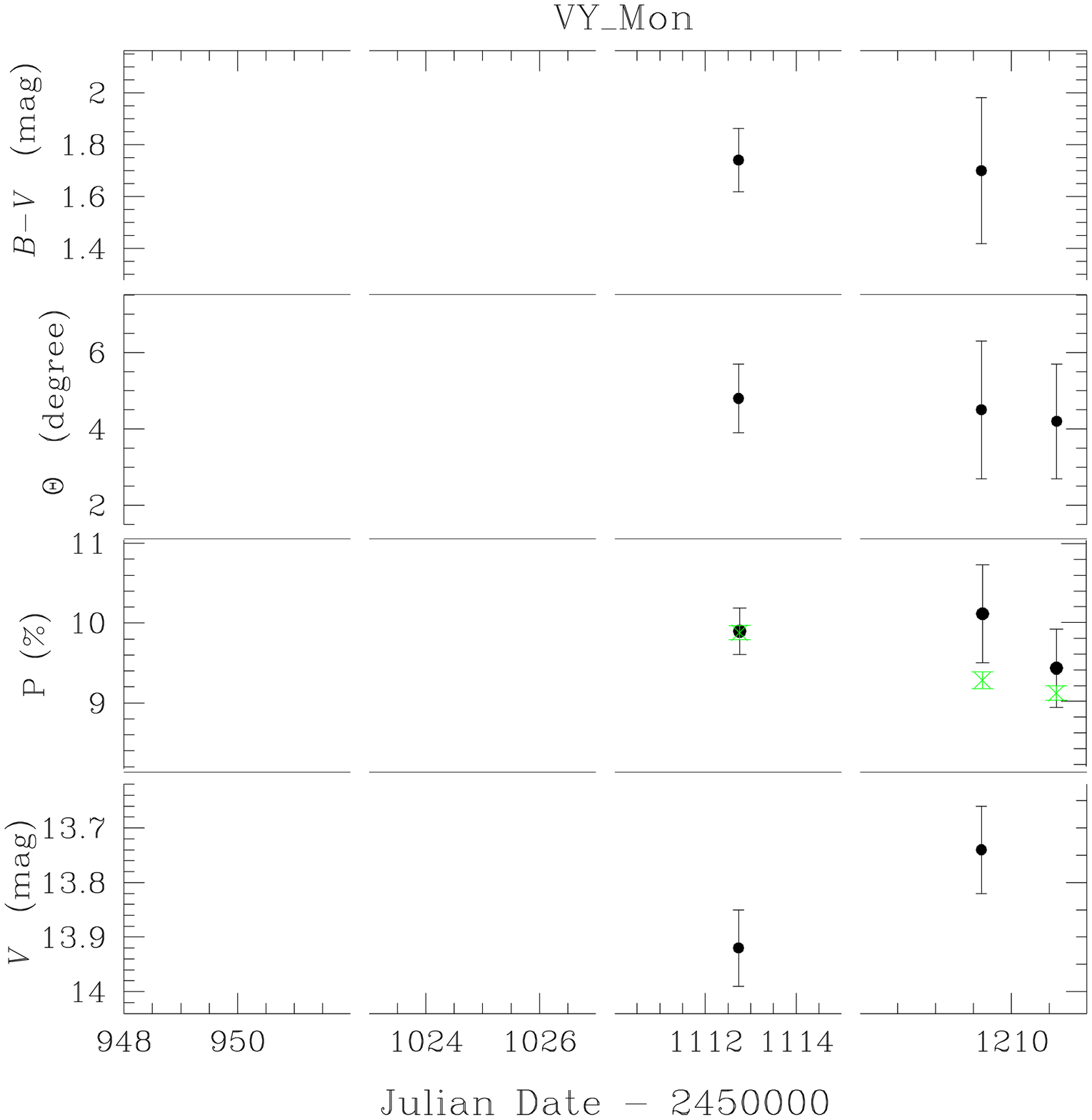}}
\mbox{\epsfxsize0.32\textwidth\epsfbox[0 125 600 685]{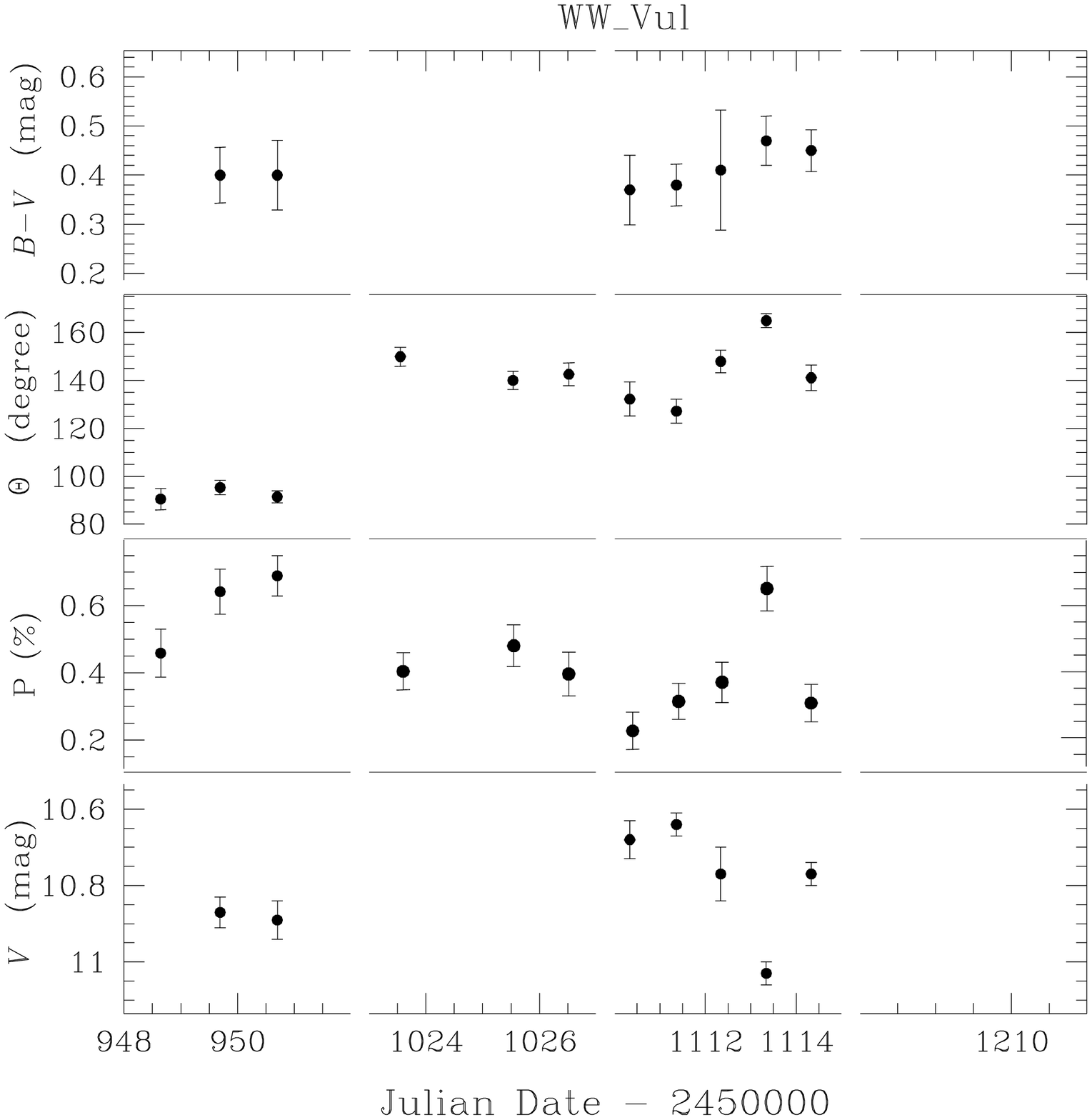}}
\caption{Objects known to exhibit the UXOR phenomenon and showing it in
our data.  Plotted are the {\it (B--V)}, Position Angle, 
Polarization and {\it V} magnitude as function of time. For selected objects (see text), the crosses  denote the {\it R} band polarization values. The four panels in Julian Date correspond to the May, July and October 1998 and January 1999 observing runs. 
\label{yesa}}
\end{figure*}
\begin{figure*}

\mbox{\epsfxsize0.32\textwidth\epsfbox[0 100 600 685]{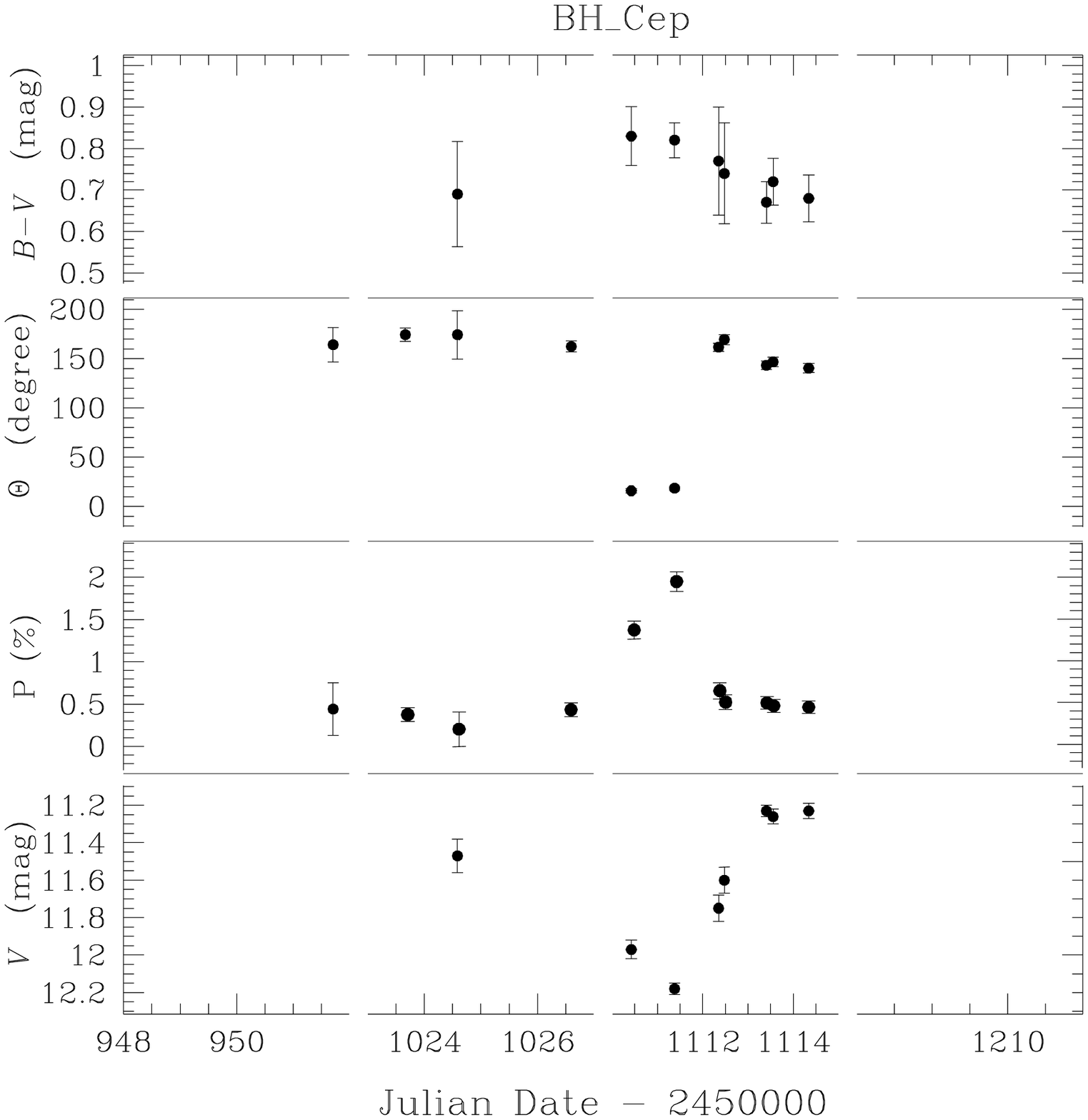}}
\mbox{\epsfxsize0.32\textwidth\epsfbox[0 100 600 685]{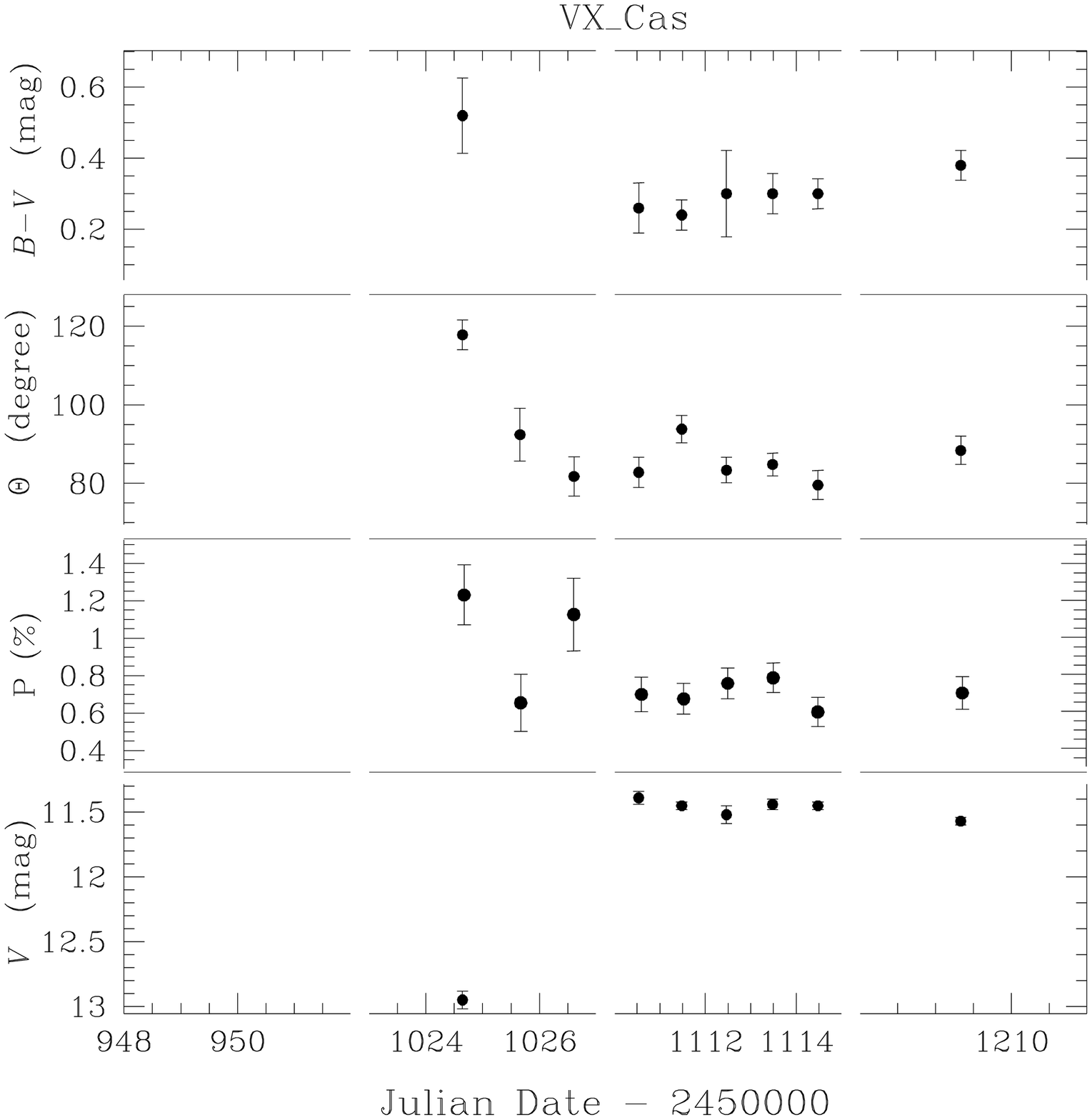}}
\mbox{\epsfxsize0.32\textwidth\epsfbox[0 100 600 685]{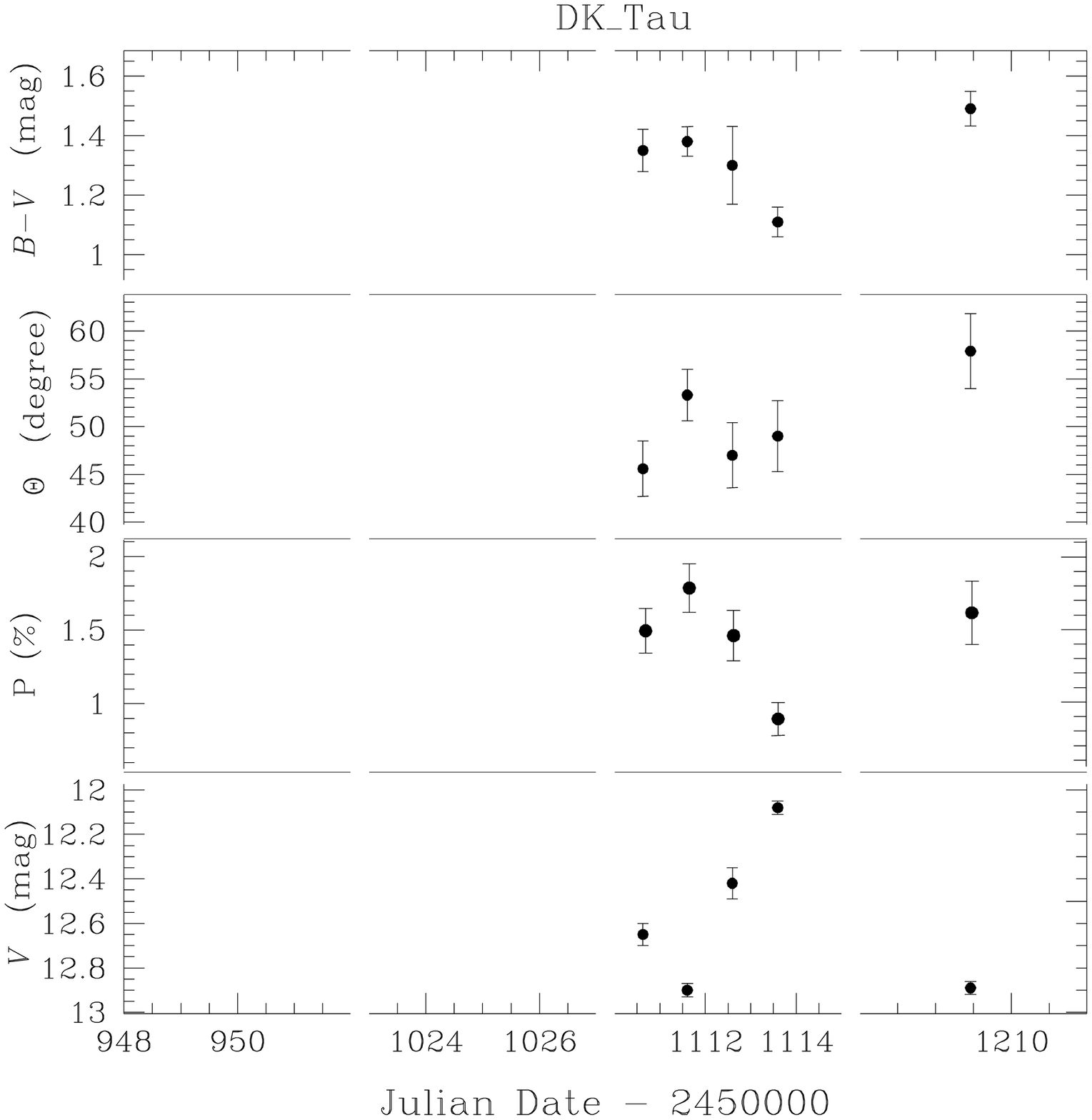}}

\mbox{\epsfxsize0.32\textwidth\epsfbox[0 125 600 685]{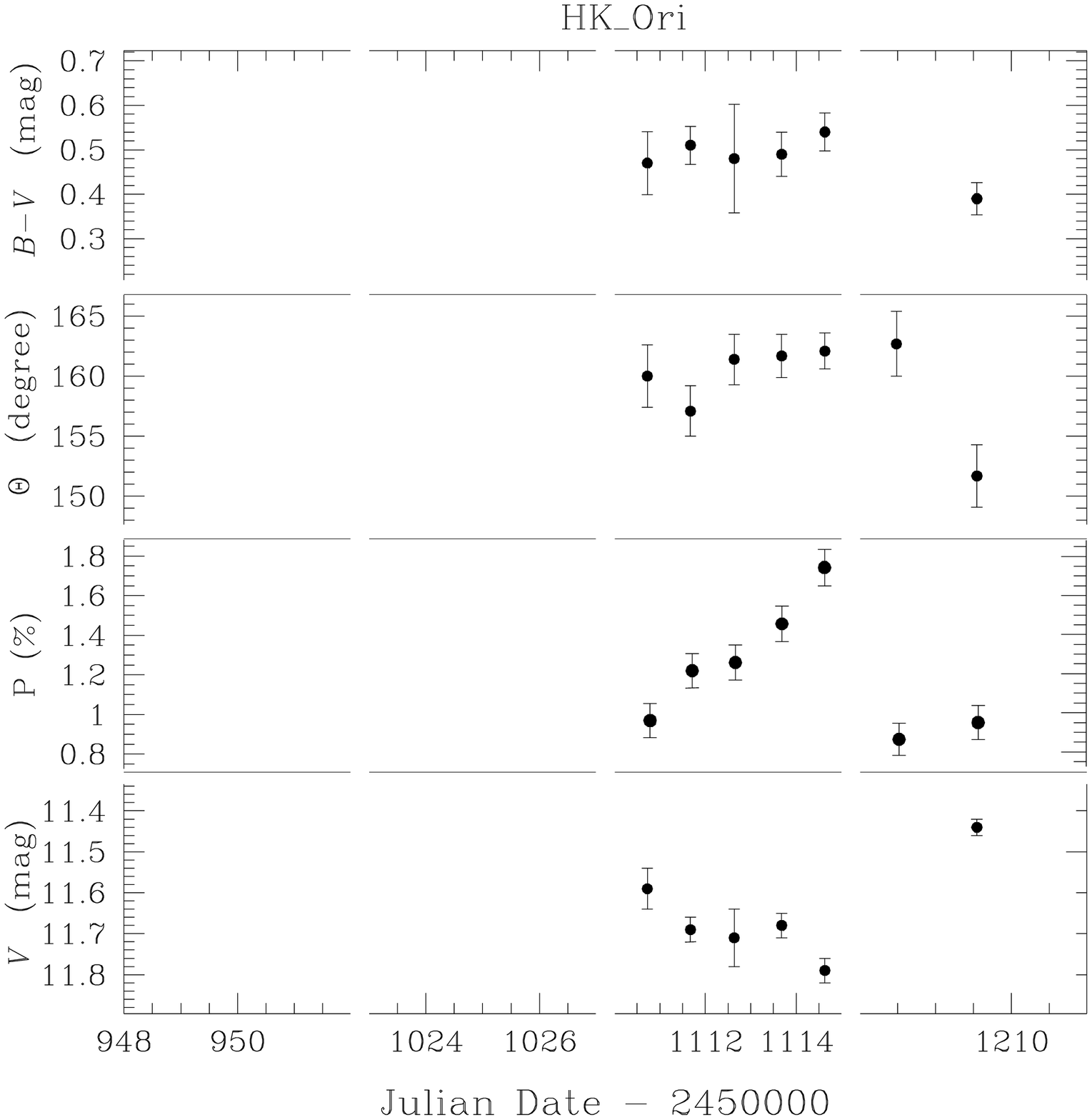}}
\mbox{\epsfxsize0.32\textwidth\epsfbox[0 125 600 685]{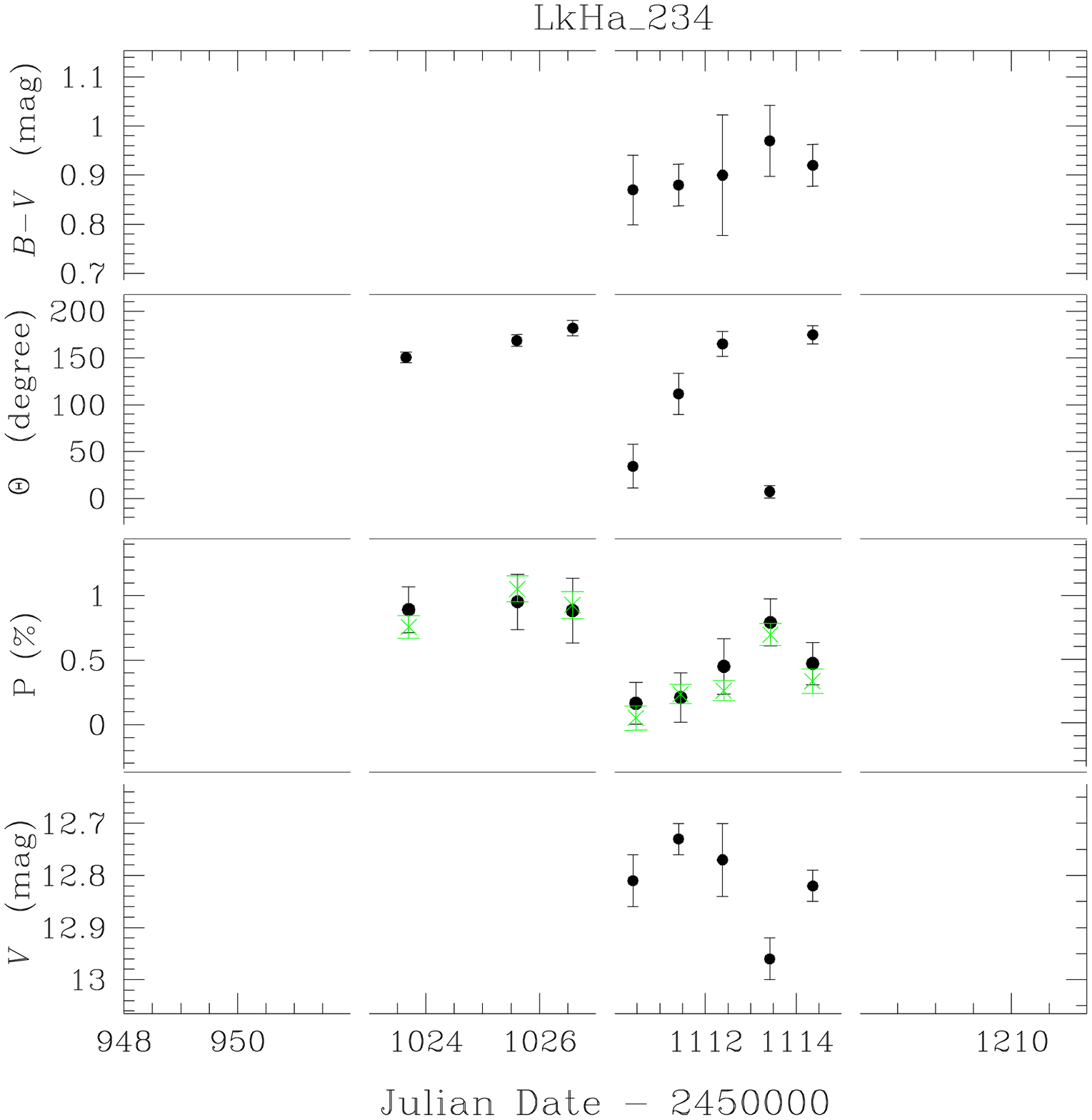}}
\mbox{\epsfxsize0.32\textwidth\epsfbox[0 125 600 685]{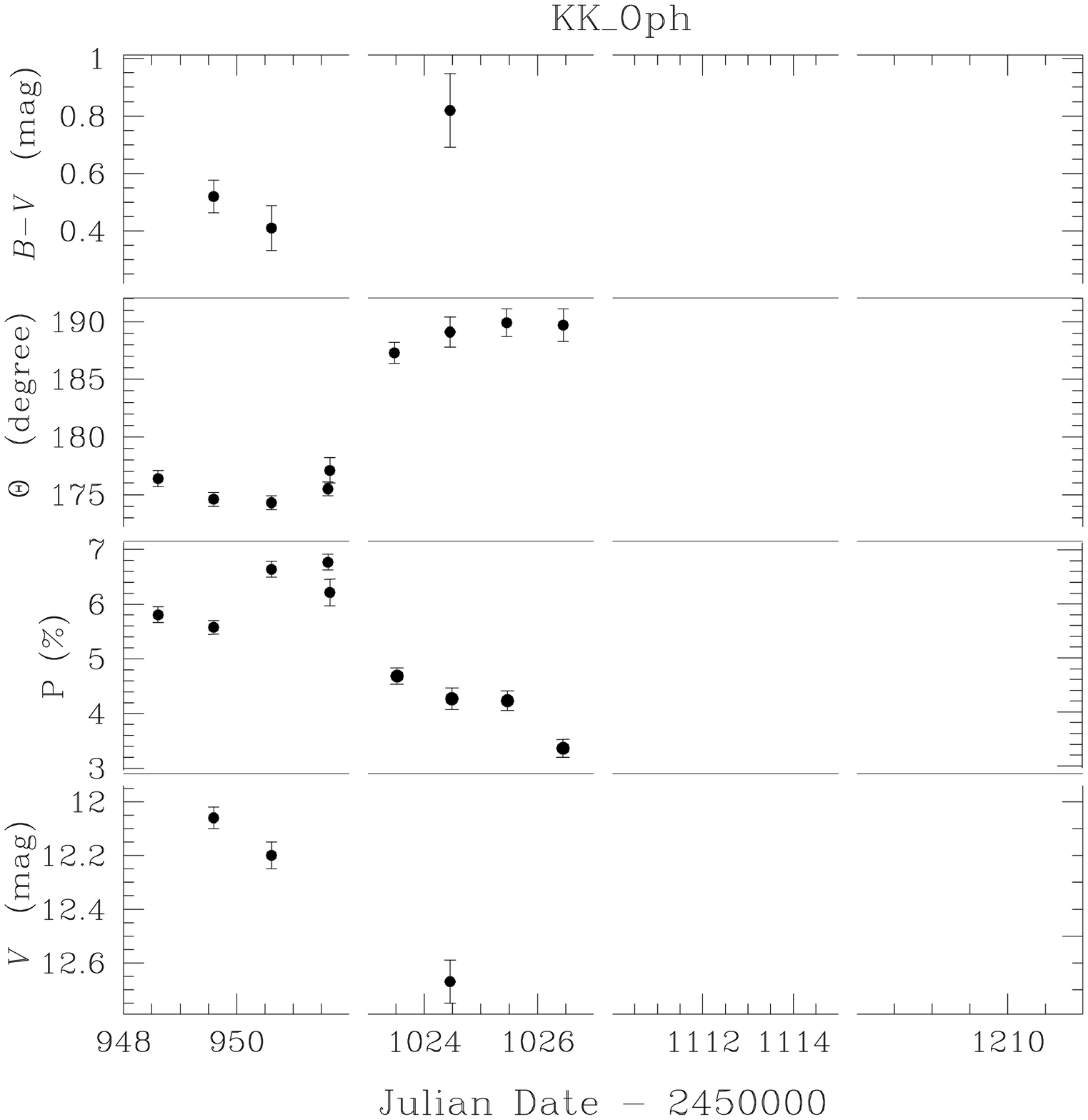}}

\mbox{\epsfxsize0.32\textwidth\epsfbox[0 150 600 685]{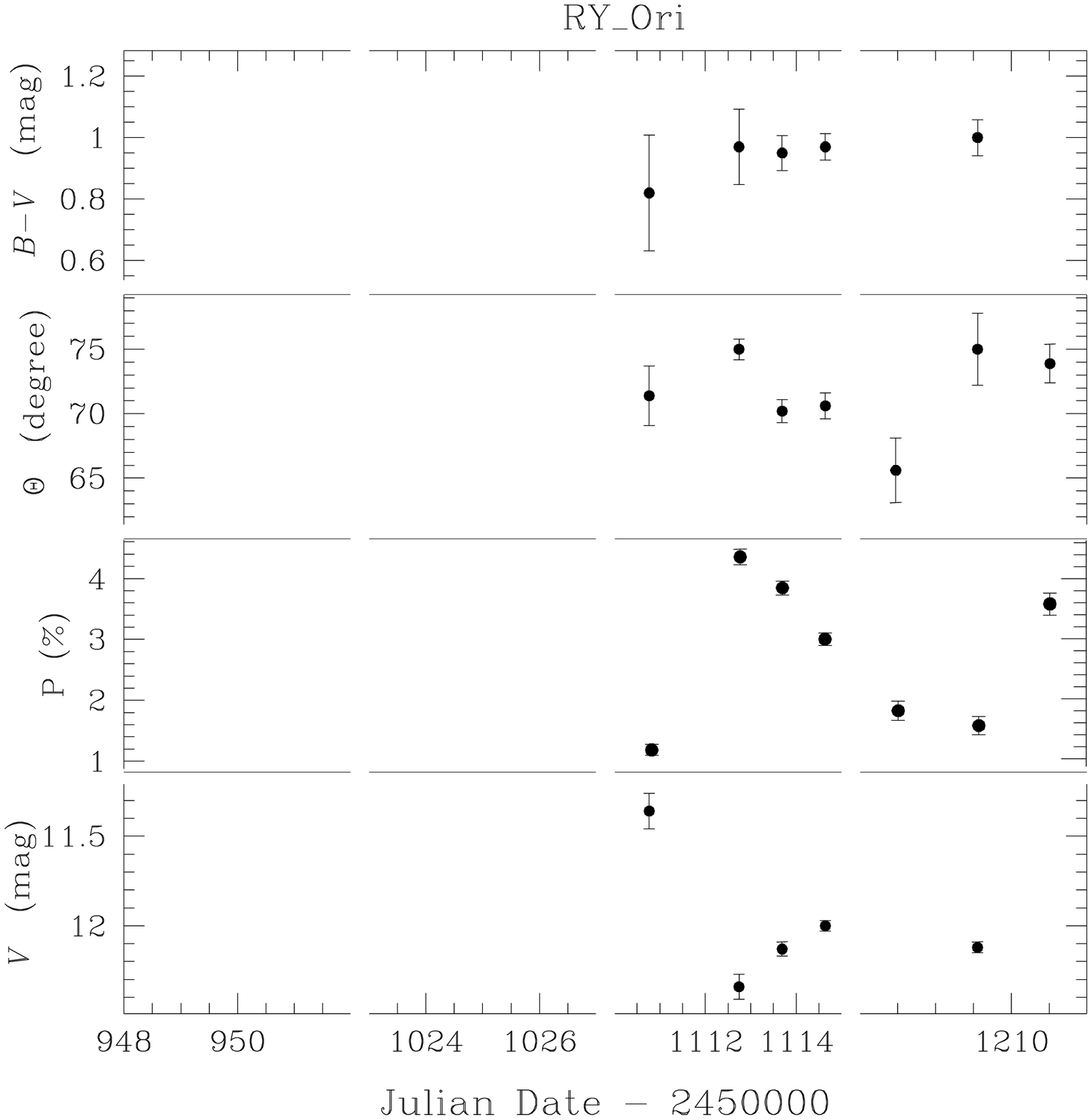}}
\caption{As the previous figure, but now for objects not previously 
known to show the UXOR behaviour.
\label{yesb}}
\end{figure*}
\end{subfigures}

\subsection{Correlation photometry - polarization}

For the objects flagged as variable in Table~\ref{log}, we investigate
the correlation between the observed polarization and photometry.
From here on, we will discuss them into two groups: those objects
showing increased polarization with fainter magnitudes (i.e. showing
the UXOR phenomenon), and those objects that do not clearly show this
behaviour in our dataset.  We base the following on the {\it V} band;
where necessary, other passbands will be discussed as well.  We point
out that the following discussion is not exhaustive as not all
observing dates were photometric, limiting the number of the combined
photo-polarimetric data.  It is also possible that our limited
time-coverage may have missed objects that are known to be variable,
but happened to be relatively stable during our observing campaigns.

\subsection{Objects showing increased polarization with 
fainter magnitudes}

In our sub-sample of 22 targets which display significant variations,
the UXOR type behaviour (in first instance, an anti-correlation
between polarization and photometry) can be found in 13, i.e. more than
half of the targets.  Their data are displayed in Fig.~\ref{yesa} and
\ref{yesb}. For six of these (BF Ori, BM And, RR Tau, VY Mon, WW Vul
and RY Tau), similar observations are present in the literature, while
for seven of these objects (BH Cep,  KK Oph, RY Ori, VX Cas, DK
Tau, HK Ori and Lk\ha 234) this is a new finding, as, to our
knowledge, no previous photo-polarimetric monitoring has been
performed for these objects.

{\it BF Ori:} The UXOR behaviour was previously discussed by Grinin et
al (1989) for BF Ori. We seem to have observed the star in a relatively bright
period, {\it V} $<$ 10 ($P$ $<$ 1\%). Grinin et al (1989) observed
minima with {\it V} $>$ 12, with $P$ reaching the 5\% level.

{\it BM And, RY Tau, and WW Vul:} For both BM And (Grinin et al, 1995)
and WW Vul (Grinin 1994) a similar behaviour as for BF Ori has been
presented in the literature. At some times, polarizations of order 6\%
were measured when the stars were at their photometric minima ({\it V}
= 14, and 12 respectively). We observed these stars in relatively
bright, longer lasting, phases.  RY Tau was observed by Petrov et al
(1999), in a somewhat brighter, lower polarization state than reported
here.

{\it RR Tau:} Although variable in the entire polarization data-set,
the variability of RR Tau is less significant in the subset containing
the photometry. The minimum in the {\it V} band photometry corresponds
to a hardly significant local maximum in polarization but is more
visible in the {\it R } band data, which are also shown for this
object.  RR Tau is known to display the UXOR phenomenon, as
Rostopchina et al (1997) has shown. They find that the {\it V} band
magnitude varies between 10 and 14, with excursions in the
polarization up to 6\%.

{\it VY Mon:} The subset containing both polarimetry and photometry of
VY Mon is limited to only two data-points.  As the errors on the {\it
V} band polarization are large, the {\it R } band data are plotted in
the figure as well. In the {\it V} band there is no polarization
variability at all, but in {\it R} and {\it I} bands the UXOR
phenomenon becomes clearly visible.  Because of the faintness and
redness of the object, the observational error bars prevent us from
detecting variability in {\it V} - a problem eased by investigating
the {\it R} band, where the UXOR phenomenon becomes visible.  Based on
two unpublished photometric data-points, Yudin \& Evans (1998) reach
the same conclusion.

We now add to this class of object an additional seven stars.
Although the photometric variability of these objects was known
already, as is evident from their names of course, it is our
simultaneous photo-polarimetric campaign that puts them in the `UXOR'
class.

{\it BH Cep:} BH Cep  shows photometric variations of 1
magnitude, coinciding with a ten-fold polarization increase from 0.2
to 2\% within two days. 

{\it VX Cas:} A slight anti-correlation is seen in the October 1998
run, the single point from July shows a deep brightness minimum
accompanied by high polarization and the reddest colour.

{\it DK Tau:} shows a 1 magnitude change and a doubling of the
polarization from 1 to 2\% within two days.

{\it  HK Ori:} displays a gradual change over a four day period by 0.4
magnitude while increasing in polarization from 1\% to 2\%.

{\it Lk\ha 234:} exhibits a drop of 0.3 magnitude in 1 day, coinciding
with an increase in polarization from 0.1 to 0.7\%.  This is more
clearly present in the {\it R} band data.

{\it KK Oph:} Only three photometric points are available for this
highly variable object. Within the May 1998 run, the two data points
show an anti-correlation between photometry and polarimetry,
indicating UXOR type behaviour. The third data point, taken in July
1998, however is the faintest photometric point, but also shows the
lowest polarization.  It may well be that the dust clouds responsible
for the obscuration in July 1998 had a different geometry (i.e. less
a-symmetric) resulting in lower polarization values and a different
inclination, as evidenced by the change in position angle between the
two runs.

{\it RY Ori:} The October 1998 run displays a very clear UXOR
behaviour, a change of 1 magnitude in brightness is accompanied by a
polarization increase of 4\%. As for the previous object, this is
not followed during the next run, where the polarization has
decreased, while the {\it V} band magnitude remained constant.

\subsection{Objects not showing 
increased polarization with  fainter magnitudes in our data}

Here, we discuss the nine stars which, from our preliminary analysis,
appear to show no anti-correlation between polarization and brightness
or for which we can make no conclusive remarks.  In some cases the
objects are mostly variable in the photometry, but the errors on the
{\it V} \ band polarization prevent us from detecting variability -
the variability being traced in the entire polarization data-set but
not in the subset containing the photometry. However, using additional
data-points, or judging from information in the literature, we find
that by and large these objects are most likely to show the UXOR
phenomenon as well. The objects that appear to show a positive
correlation between polarization and magnitude, rather than an
anti-correlation, do so when different observing epochs
are compared. This could well be explained within the UXOR scenario if
one assumes that different dust-clouds are responsible for the
respective polarization and obscuration.

{\it SV Cep:} SV Cep's polarization variability has decreased
substantially in the subset containing the photometry, which is
probably why our data do not show the UXOR phenomenon. A recent paper
by Rostopchina et al (2000) presents 10 year long monitoring of the
star, and only one deep minimum was detected. Indeed.  the
polarization increased during this event, suggesting SV Cep is also a
member of the UXOR class. It is plausible that the rareness of the
polarimetric events is the reason why we missed the effect altogether.

{\it V350 Ori:} The subset of V350 Ori containing the photometry with
only three data points, does show significant polarization
variations. Here, the observational errors on the photometry make an
assessment of the variations inconclusive.  Interpreting their own,
sparse, data of the object, Yudin \& Evans (1998) claim that it shows
the UXOR-type behaviour, but it is unclear whether this conclusion is
based on simultaneously obtained photometry.

\begin{figure*}
\mbox{\epsfxsize0.32\textwidth\epsfbox[0 100 600 685]{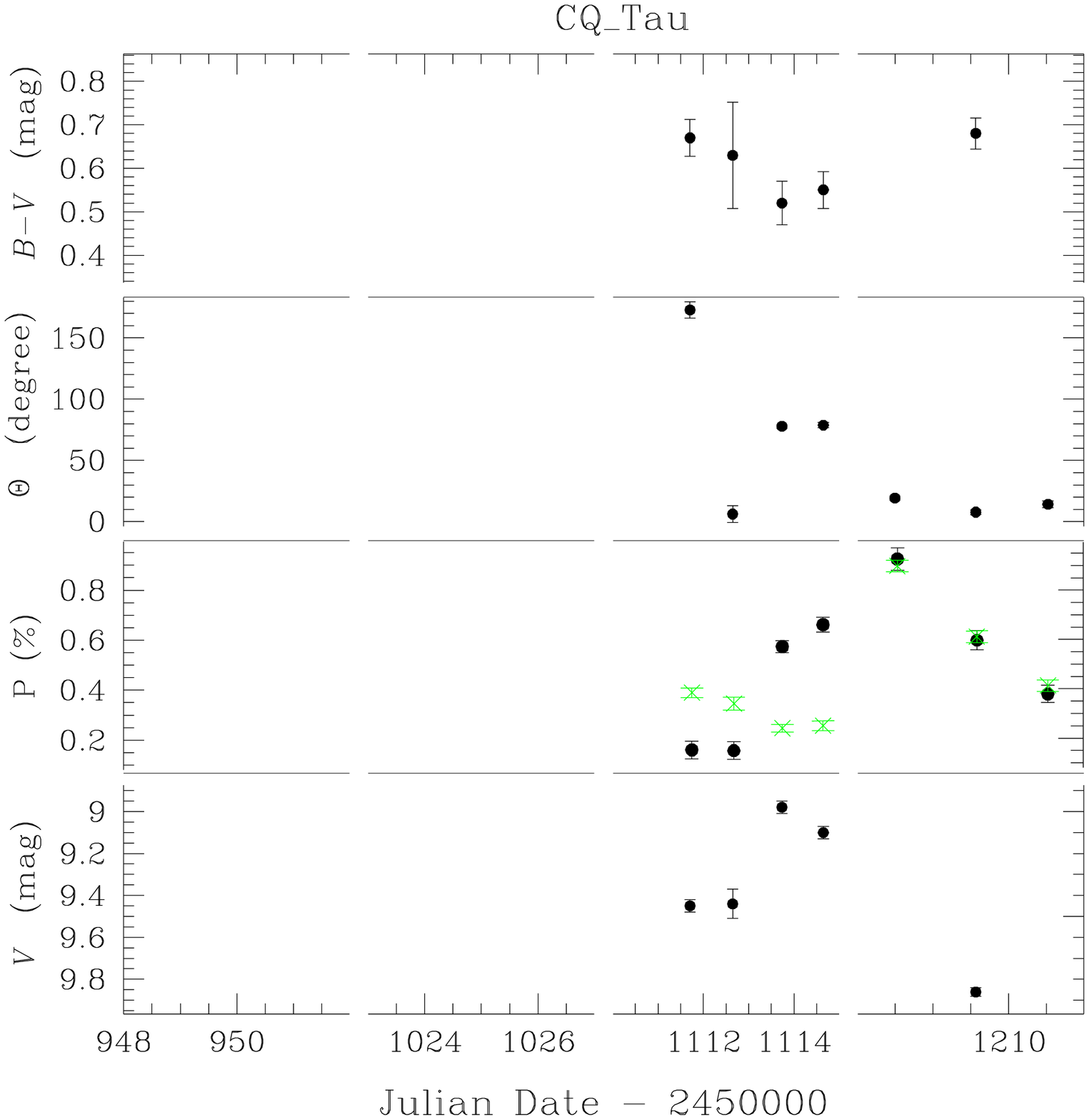}}
\mbox{\epsfxsize0.32\textwidth\epsfbox[0 100 600 685]{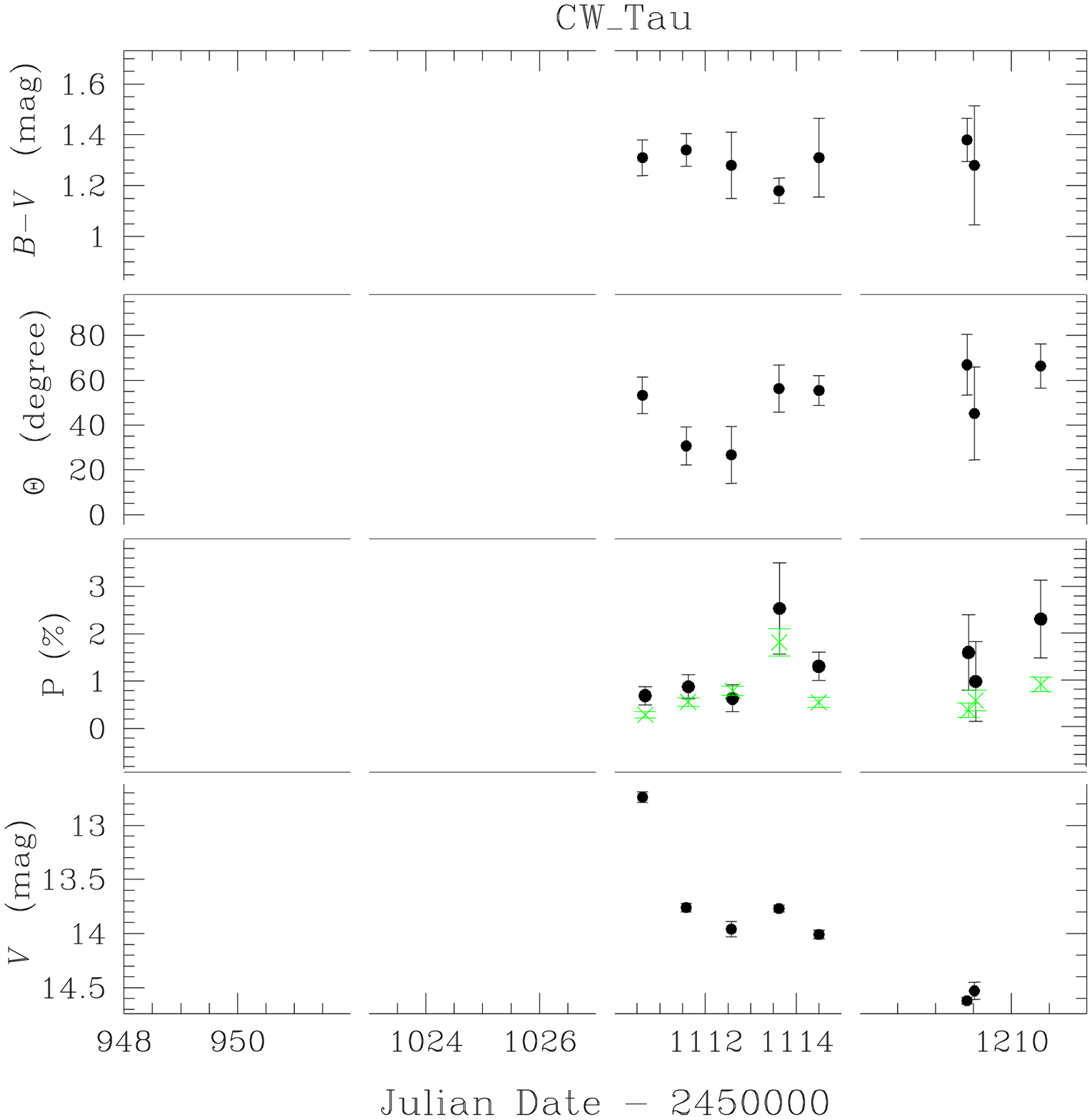}}
\mbox{\epsfxsize0.32\textwidth\epsfbox[0 100 600 685]{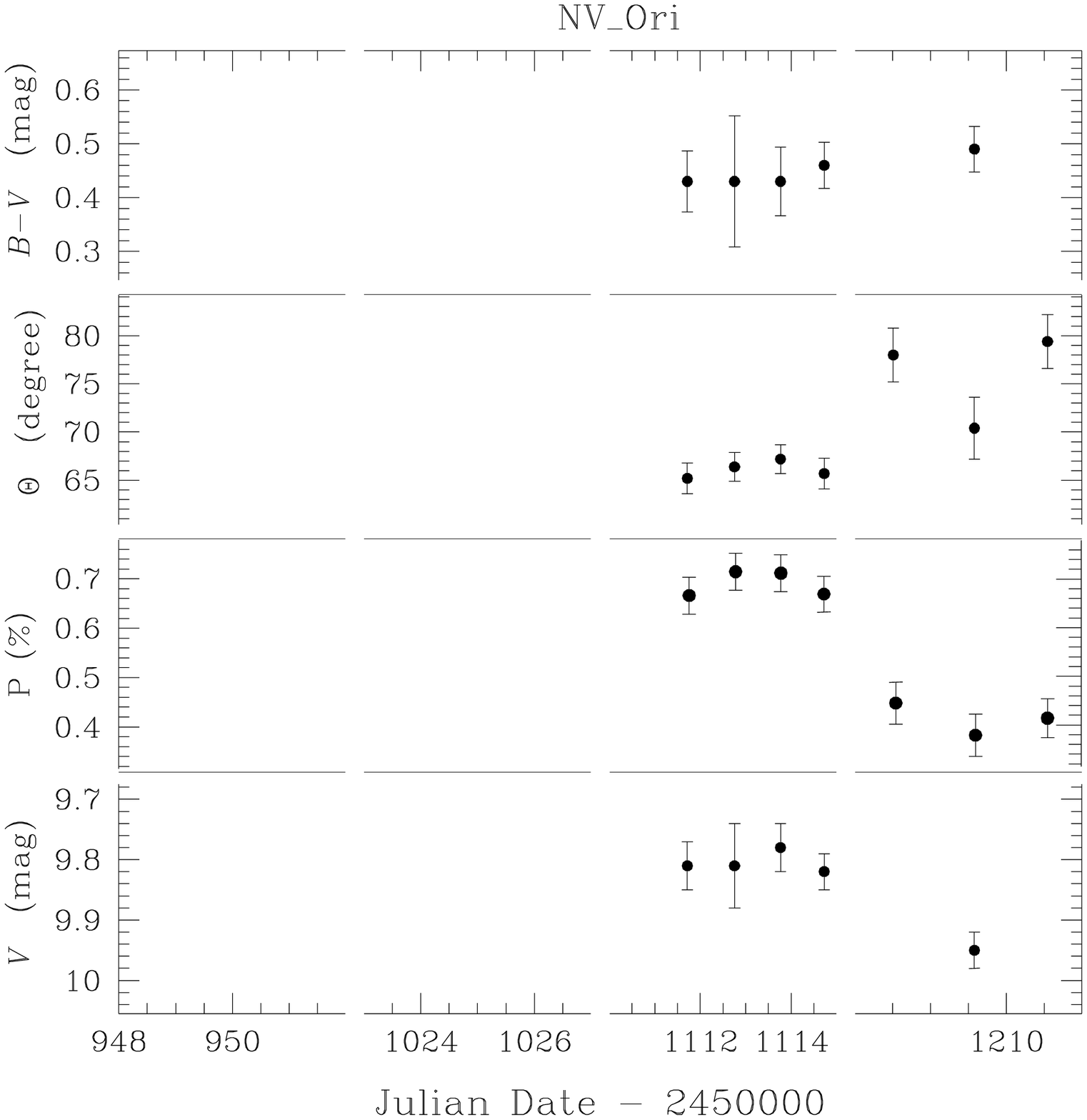}}

\mbox{\epsfxsize0.32\textwidth\epsfbox[0 125 600 685]{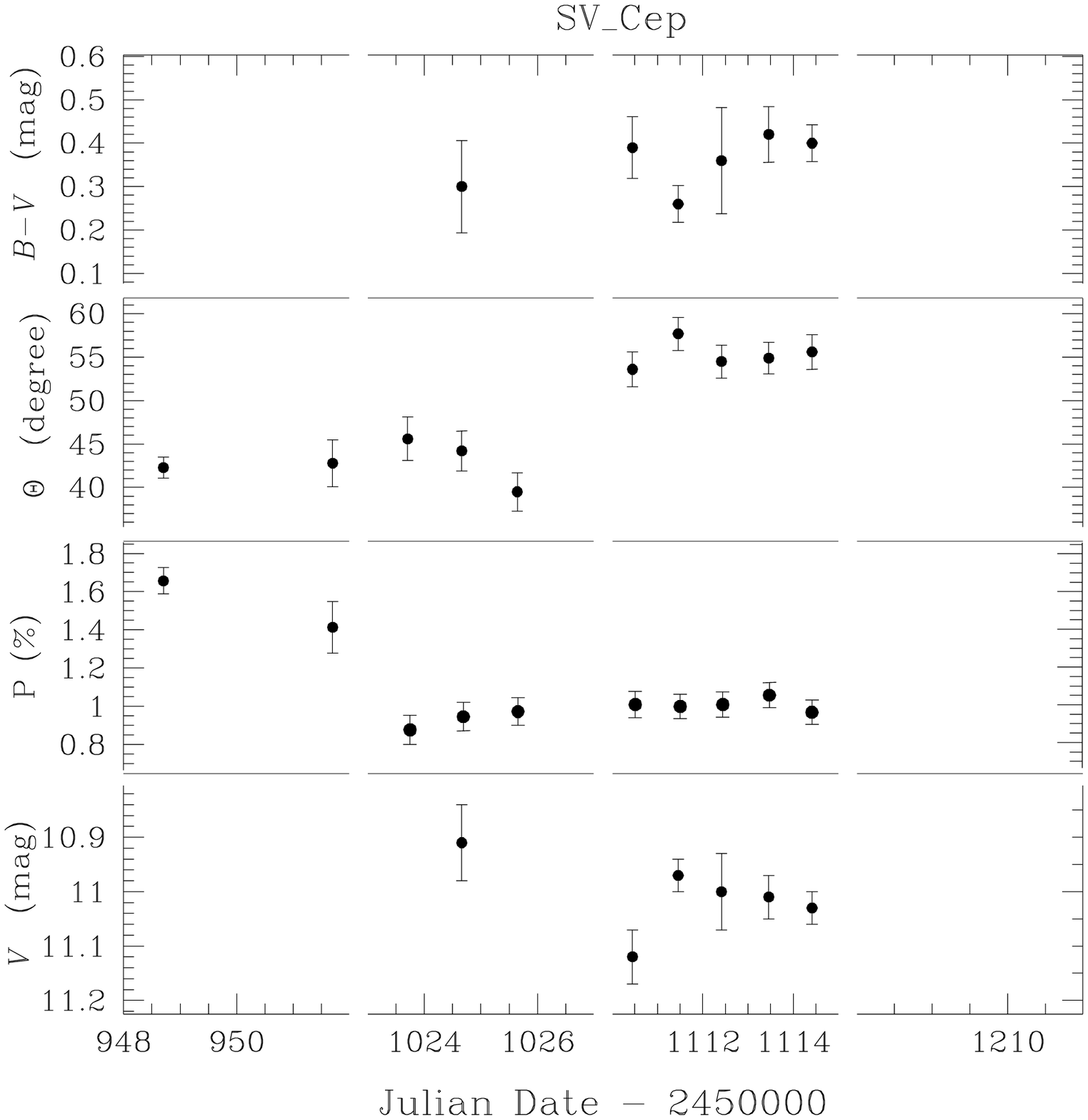}}
\mbox{\epsfxsize0.32\textwidth\epsfbox[0 125 600 685]{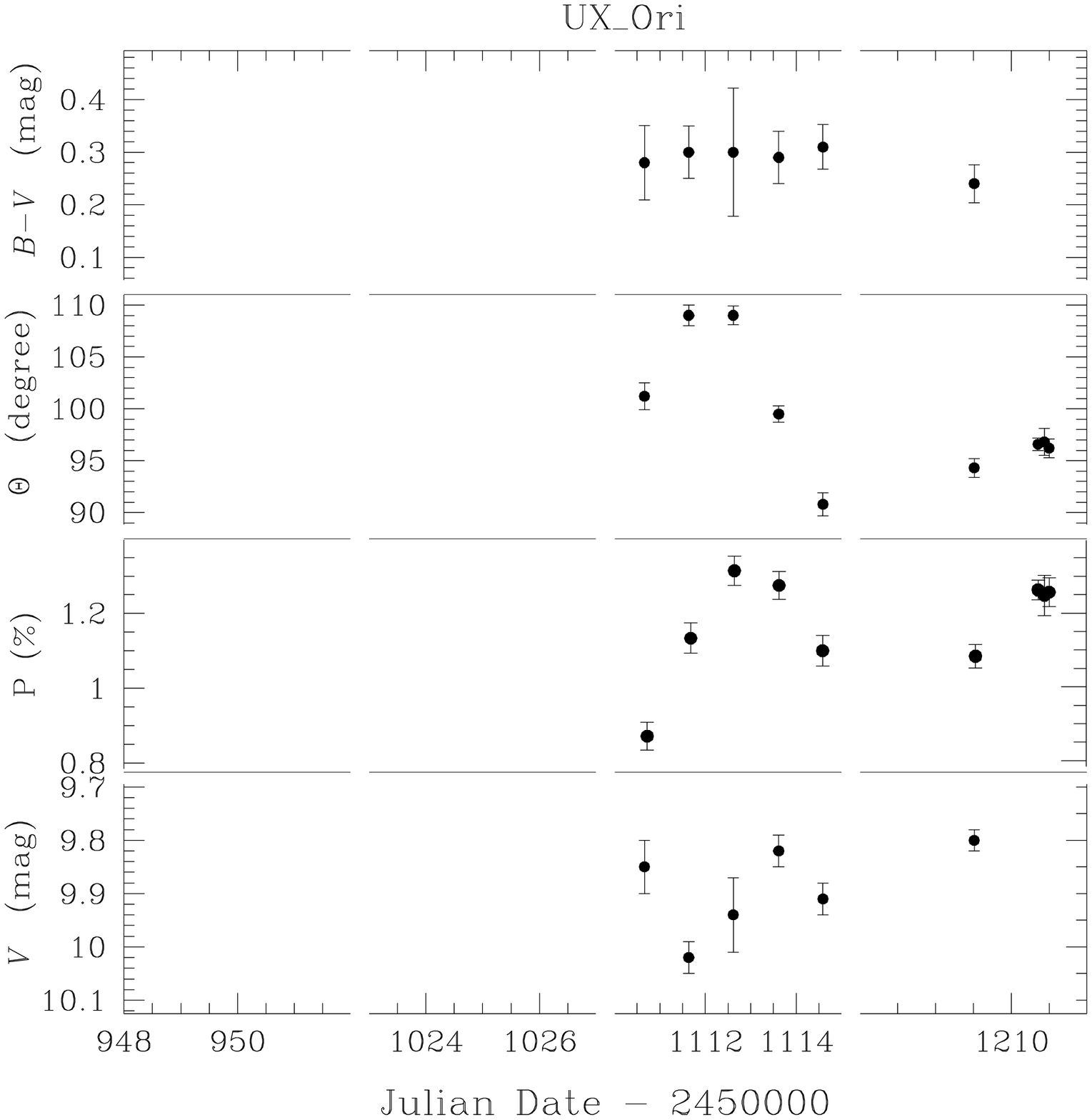}}
\mbox{\epsfxsize0.32\textwidth\epsfbox[0 125 600 685]{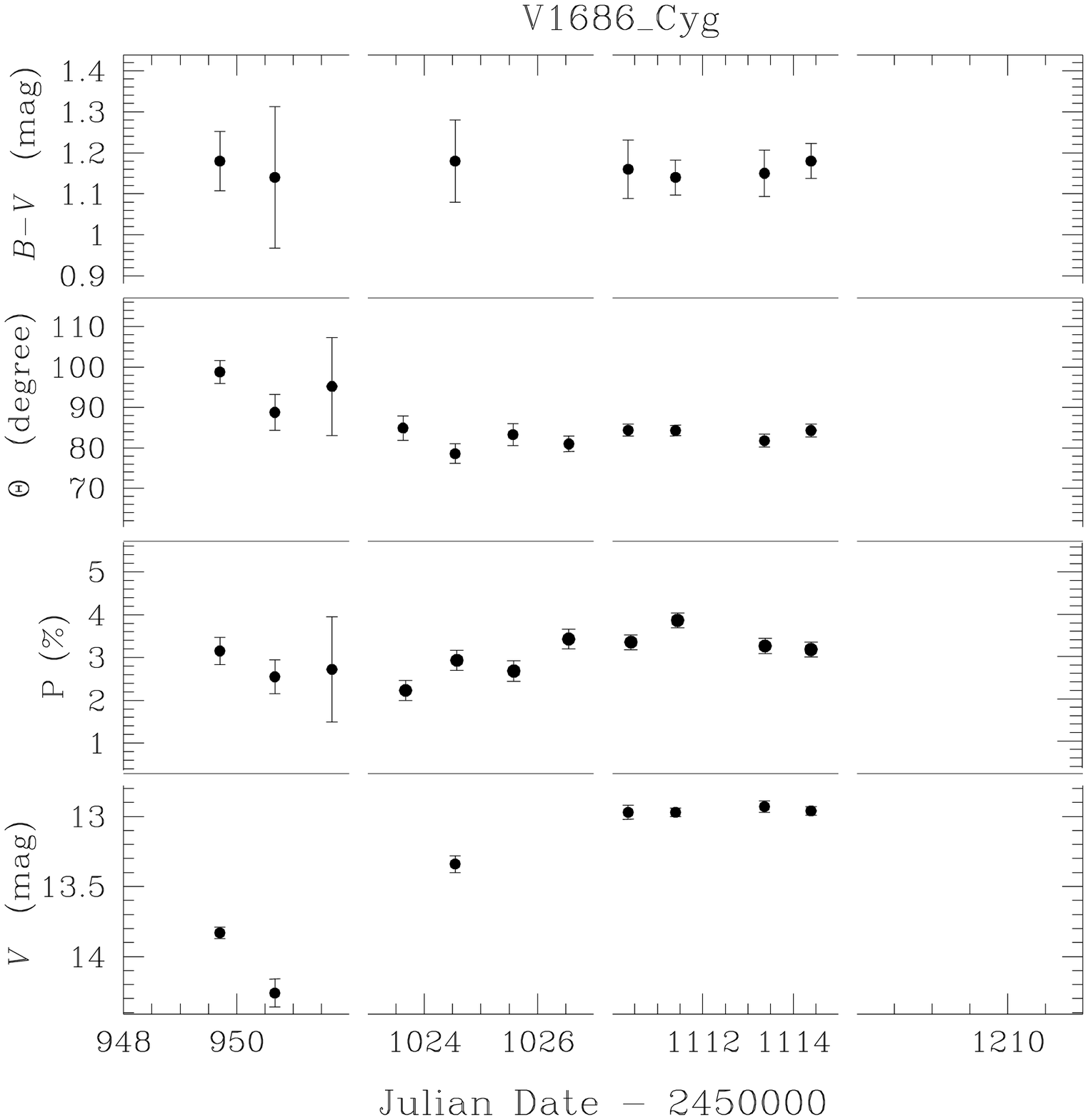}}

\mbox{\epsfxsize0.32\textwidth\epsfbox[0 150 600 685]{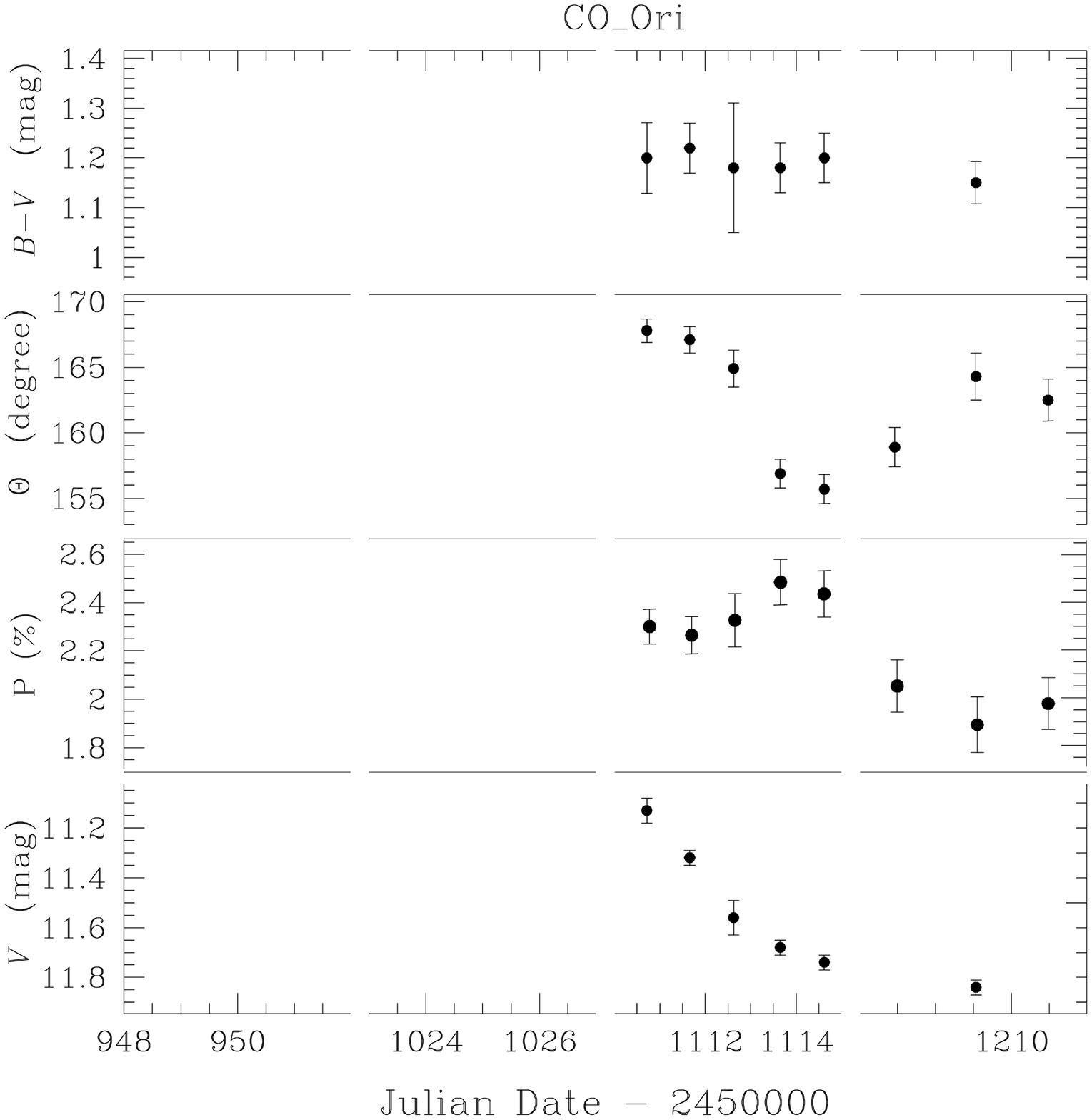}}
\mbox{\epsfxsize0.32\textwidth\epsfbox[0 150 600 685]{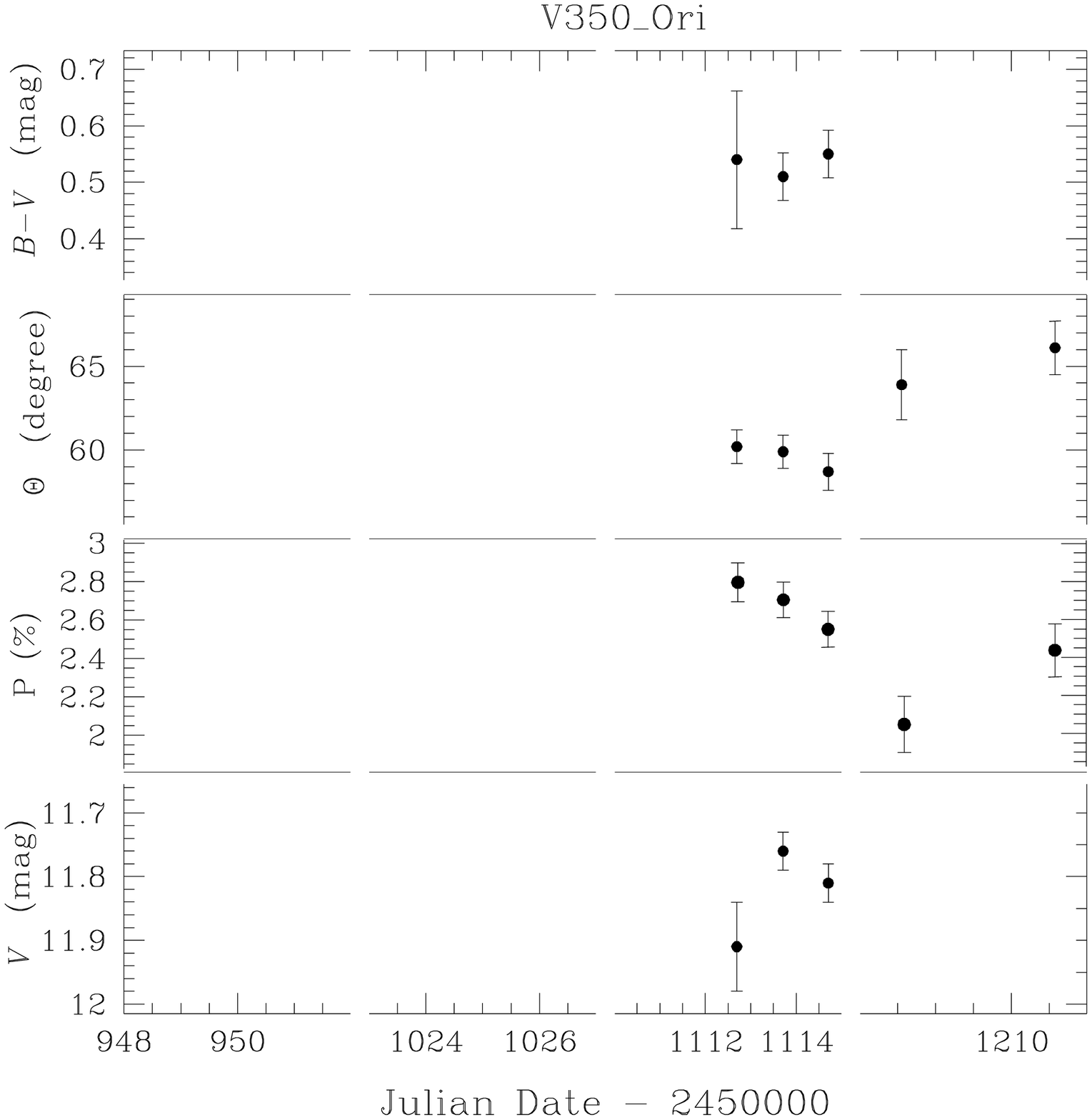}}
\mbox{\epsfxsize0.32\textwidth\epsfbox[0 150 600 685]{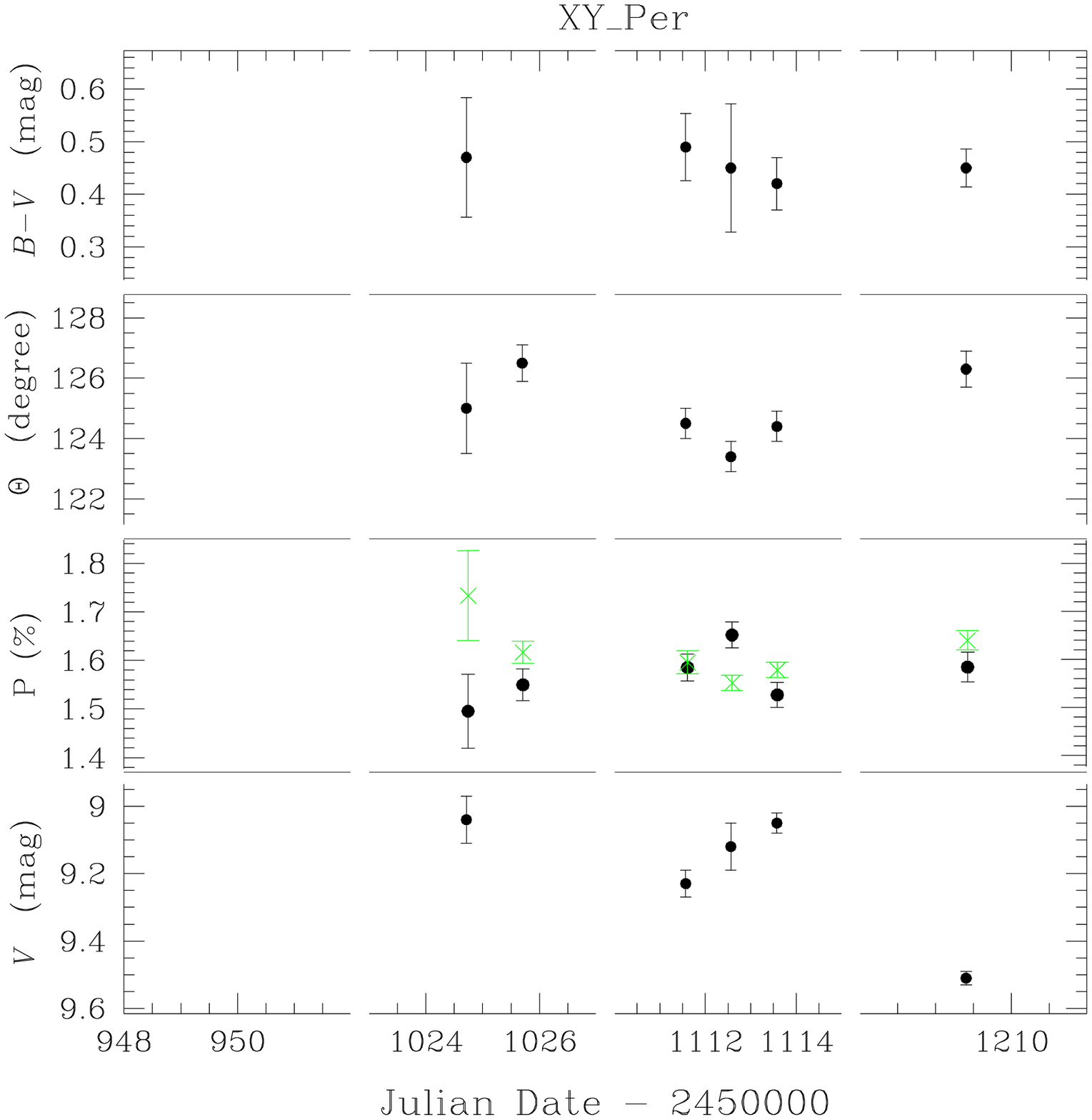}}
\caption{As the previous figures, but now for 
objects that do not seem to exhibit the UXOR behaviour. 
The open squares in the figures for VY Mon and XY Per denote
the {\it I } band data.
\label{maybe}}
\end{figure*}

{\it CO Ori: } The photometric data of CO Ori show a gradual decrease
in brightness over five days in October 1998, while the polarization data
show a slight, but not significant, increase in polarization.  When
the star is at its faintest detected level, it exhibits the lowest
polarization -contrary to the usual UXOR behaviour.  This `deviating'
point is from another epoch (January 1999), a period when the observed
polarization was consistently lower than in October 1998, but with
only one photometric night available. As for KK Oph and RY Ori
discussed above, it may well be that the dust clouds responsible for
the obscuration in January 1999 had a different geometry If true, then
CO Ori is another object to be added to the UXOR category.

{\it NV Ori: } During the October 1998 run, the data of NV Ori cluster
within the errors bars around {\it V} $\sim$ 9.8, while in
polarization small changes are detected. As for CO Ori, in the
subsequent observing epoch, both the polarization and optical
brightness have decreased dramatically, If the object was obscured,
but relatively quiescent in October, and a different cloud is causing
the obscuration in January 1999, it is well possible that NV Ori is a
new member of the UXOR class.

{\it CQ Tau: } As can be seen in Fig.~\ref{maybe}, CQ Tau becomes redder with
fainter {\it V} band magnitudes, but the polarization is a minimum in
the middle of the photometric range. An effect also seen in the {\it
B} band (not shown). A clue to CQ Tau's behaviour is provided by the
fact that we can see that the {\it R } (shown in the figure) and {\it
I} band polarization behave exactly as expected from an UXOR variable.

It is interesting to note that the change in polarization is more
clearly detectable at longer wavelengths. Although this is
counter-intuitive, since polarization increases with shorter
wavelengths, this is not a new result. Inspection of the figures in
e.g. Grinin et al (1991), shows that this indeed
happens in some cases. The possible reason is that we are looking at the
polarization vector addition of the circumstellar polarization and
interstellar polarization. Since both can have a different wavelength
dependence, the (vector-)addition of the two will show a different
wavelength dependence. The change in PA from the first two
days in October to the latter two is dramatic. In the case of one of
the two components varying, we will then also see a varying total
component. In principle, in the photometric minimum of the star, the
circumstellar polarization should dominate, while in photometric
maximum the interstellar component, depending on its magnitude, can
dominate.

Grinin (1994) pointed out that CQ Tau does show the UXOR phenomenon,
but with some exceptions, sometimes when the star has larger
polarization, it is brighter.

{\it CW Tau: } For convenience, the {\it R} band data are also shown
for this object, as the faintness of the star prevents us from discussing
the lower quality {\it V} band data.  CW Tau shows photometric
variations in excess of 1.5 mag, but, with the exception of one
higher polarization point in both {\it V} and {\it R} remains quite
constant in polarization. It was especially the changes in Position
Angle that were picked up by the variability-software, and we only
note at this stage that the interplay between circumstellar and
interstellar polarization may play a role for this object. Clearly, CW
Tau deserves further study.

{\it UX Ori:} It is ironic  that UX Orionis should be
classified  as not displaying the UXOR behaviour. There is no
obvious trend visible in the data, but the range of {\it V} band
photometry is the smallest observed in the sample of polarimetrically
variable objects, with variations less than 0.25 mag. Comparison with
data of Grinin et al (1994) reveals that we have caught the object in
a bright, low polarization spell. The changes in polarization are
accompanied with significant changes in PA, indicating that several
polarization mechanisms may be acting on the object. We previously
discussed the interplay between circumstellar dust polarization and
the interstellar polarization, however, we point out that low
amplitude photometric variations do not have to necessarily be due to
the dust obscuration phenomenon alone. 
As shown for example by Oudmaijer
and Drew (1999), electron-scattering can contribute up to 1\% to the
total polarization towards Herbig Be stars. If the excitation
conditions within the ionized envelope change, 
slight changes in polarization can be readily detected.

{\it XY Per and V1686 Cyg:} Despite the comparatively large data
coverage for these two stars, we are not in a position to classify the
stars' behaviour during our observations.  Although the stars show
photometric variations of more than 1.5 magnitude, and relatively
strong polarization variability it is hard to describe a trend.

Of the nine objects discussed in this section, we find that four
objects (CQ Tau, UX Ori, SV Cep, V350 Ori) had previously been
reported to be UXOR variable stars. These stars do not show the
expected anti-correlation between polarimetry and photometry in our
data. The reason is most likely that our subset containing the
photometric data was too small to find this behaviour as well.

For two objects (CO Ori, NV Ori) we find opposed behaviour to the
expected anti-correlation. Their behaviour can be incorporated into
the explanation of the UXOR phenomenon, provided different dust clouds
act on the polarization, as was found for example in the case of KK
Oph.

For the remaining three objects (CW Tau, V1686 Cyg and XY Per), we
are not able to classify their behaviour, and further studies are
warranted.

\subsection{Conclusion on the polarization variability}

From the 22 objects in our data-set that showed significant
variability, most of them appear to display the UXOR phenomenon
i.e. show increased polarization when fainter. This is a rather loose
definition of the UXOR phenomenon, as the colour changes play a
central role in the explanation of the phenomenon as well. However, on
the whole, we have not detected the strong photometric changes of more
than 4 magnitudes with their associated colour changes of order
0.3-0.5 magnitudes. This is most likely due to the fact that more
often than not the UXOR variables are in their bright state,
preventing us from detecting their more extreme behaviour in our
comparatively short monitoring programme.

Some objects, which are variable in their entire polarization dataset,
were not variable in the subset containing the photometry. In
those cases, existing literature on these objects showed that they are
UXOR variables.  

Indeed, most of the objects known as UXOR variable, were recovered via
our statistical tests. A notable exception is VV Ser (Kardopolov et al
1991) in which variations are observed, but not in such a way that
they satisfy our criteria outlined in Sect. 3. It is most likely our
comparatively sparse sampling that prevents us from catching all such
objects, but on the whole the detection rate has been very high.

We add a total of seven new objects to the class of UXORs, while, if we
were to include NV Ori and CO Ori, showing photo-polarimetric
variability as well, we increase the sample by nine.  These objects were
previously not recognized as such because they had not been
photo-polarimetrically monitored before.

\section{Polarization properties of the Main Sequence objects}

\begin{table*}
\begin{center}
\caption{
The Vega-type and related objects in our sample. Spectral types are taken from Mora et al (2001) or from SIMBAD (marked with `$^S$'). The photometry is taken from this paper, of from SIMBAD (marked with `$^S$'). The polarization values are weighted means, also from this paper. $A_V$ is derived from the spectral types and photometry, and the intrinsic {\it (B -- V)} values listed in Schmidt-Kaler (1982). $\lambda_{\rm max}, K$, and $\chi^2_{\rm red}$ result from fitting Serkowski laws through the data (see text).
The `types' are, unless otherwise noted, Vega-type.
\label{vegas}}
\begin{tabular}{llllrlrlllll}
\hline
\hline
Name  & Spectral & {\it B} & {\it V} & $P_V$ & $\sigma_{P_V}$ & $A_V$  & $\lambda_{\rm max}$ & $K$ & $\chi^2_{\rm red}$&  type\\
      & Type     &($\pm$0.05)&($\pm$0.05)&(\%)&  (\%)         & ($\pm$0.2)& ($\mu$m) \\
\hline
   17 Sex &     A0V&    5.94&  5.92  &   $<$0.10 &      & 0.12 &   &  &    &A-shell\\
   24 Cvn &     A4V&    4.81&  4.66  &   $<$0.05 &      & 0.12 &   &  &    &A-shell\\
   49 Cet &     A4V&5.68$^S$&5.62$^S$&  $<$0.18  &      &--0.16&   &  &    &\\
   51 Oph &B9.5IIIe&    4.82&  4.83  &  0.47     & 0.02 &  0.12&0.35 & 1.15  & 4.8   & Herbig Ae/Be/Vega\\
 BD31 643 &     B5V&    9.27&  8.56  &   1.27    & 0.03 &  2.73&0.80 & 0.95  & 0.6   &\\
BD31 643C &A3IV$^S$&   11.18&  10.3  &  0.70     & 0.03 &  2.45&0.90 & 0.15  & 3.5   &\\
 HD 23362 &   K5III&    9.72&   8.1  &  0.66     & 0.04 & 0.37 &0.48 & 0.57 & 0.85   &\\
 HD 23680 &    G5IV&    9.84&  8.46  &   1.18    & 0.06 & 1.89 &0.72 & 0.62 & 0.49   &\\
 HD 34700 &    G0IV&    9.73&  9.14  &  0.35     & 0.06 &--0.06&0.30 & 0.50 & 1.13   &\\
 HD 58647 &   B9IV:&    6.92&  6.85  &  0.16     & 0.02 & 0.43 &0.30 & 0.02 & 2.11   &H/ZAMS\\
HD 109085 &    F2V &    4.66&  4.32  &  $<$0.12  &      &--0.03&   &  &    &\\
HD 123160 &   K5III&   10.22&  8.69  &  0.30     & 0.04 & 0.09 &0.55 & 1.52 & 0.08   &\\
HD 141569 &     A0V&     7.2&  7.11  &  0.62     & 0.02 & 0.34 &0.65 & 1.22 & 1.25   &H/ZAMS\\
HD 142666 &     A8V&    9.16&  8.68  &  0.71     & 0.04 & 0.71 &0.90 & 0.32 & 0.33   &H/ZAMS\\
HD 142764 &     K7V&   11.32&  9.56  &   1.56    & 0.03 & 1.33 &0.50 & 1.05 & 0.07   &\\
HD 163296 &     A1V&    6.92&  6.86  &  $<$0.04  & 0.02 &  0.16&   &  &    & H/ZAMS\\
HD 199143 &     F6V&    7.77&  7.23  & $<$0.13   &      &  0.19&   &  &    &H/ZAMS\\
HD 233517 &     K0V&   11.03&  9.69  &   1.64    & 0.03 &  1.64&0.32 & 0.18 & 2.02   &\\
    HR 10 &     A0V&    6.36&  6.24  & $<$0.12   &      &  0.43&   &  &    & A-shell\\
    HR 26 &     B9V&    5.48&  5.45  & $<$0.17   &      &  0.31&   &  &    &PTT/MS \\
   HR 419 & B9V$^S$&     6.6&  6.65  &  0.19     & 0.06 &  0.06&   &  &    &PTT \\
  HR 1369 & B9V$^S$&    5.31&  5.38  & $<$0.15   &      &  0.0 &   &  &    &PTT \\
 HR 1847A &     B5V&    6.12&  6.11  &  0.63     & 0.04 &  0.56&0.58 & 0.85 & 0.51   &PTT/MS\\
 HR 1847B &B7IV$^S$&    6.52&  6.48  &  0.63     & 0.02 &  0.53&0.64 & 0.62 & 0.18   &PTT\\
  HR 2174 &    A2V:&     5.8&  5.73  & $<$0.10   &      &  0.06&   &  &    &PTT/MS\\
 HR 2174B &  A0$^S$& 6.9$^S$&6.85$^S$&    $<$0.13&      &  0.22&   &  &    &PTT\\
 HR 4757B &     K2V&9.46$^S$&8.51$^S$&    $<$0.27&      &  0.12&   &  &    &PTT\\
 HR 5422A &     A0V&    6.04&  6.09  & 0.08      & 0.02 &--0.09&0.59 &1.98  & 0.12   &PTT/MS\\
 HR 5422B &     K0V&   11.93& 11.37  & $<$0.13   &      &--0.78&   &  &    &PTT \\
  HR 9043 &     A5V&    6.44&  6.27  & $<$0.12   &      &  0.06&   &  &    &\\
$\lambda$ Boo & A1V&    4.26&  4.22  & $<$0.04   &      &  0.09&   &  &    &\\
\hline
\hline
\end{tabular}
\end{center}
\end{table*}

In this section, we discuss the polarization properties of the main
sequence stars in our sample. The sample comprises Vega-type stars, A
shell type stars and binary systems consisting of an early type main sequence
primary and a late secondary classified as post-T Tauri star. In some
of the latter cases, only the primary was observed. We further
included some Herbig Ae/Be stars that are known to be located very
close to the main sequence and 51 Oph that is sometimes classified as
a Herbig Ae/Be star and sometimes listed as Vega-type.

For many of these stars, listed in Table~\ref{vegas}, we present the
first polarization measurements, and for virtually all of these stars
photo-polarimetric monitoring has been performed for the first time.

The main question that we will address here is  whether
these objects have intrinsic polarization due to the presence of a
stellar disk.  As mentioned in the Introduction only observed
variability in the polarization of an object presents strong evidence
for the presence of intrinsic polarization. However, as shown above,
only the PMS stars show a high degree of variability, hence we will
try to answer the question of the presence of intrinsic polarization
via different routes. Not all of these methods are conclusive, or as
strong as the variability-test, but they will at least provide us with
a sample of interest.

\subsection{The relation between observed polarization and extinction}

\begin{figure}
\mbox{\epsfxsize0.5\textwidth\epsfbox{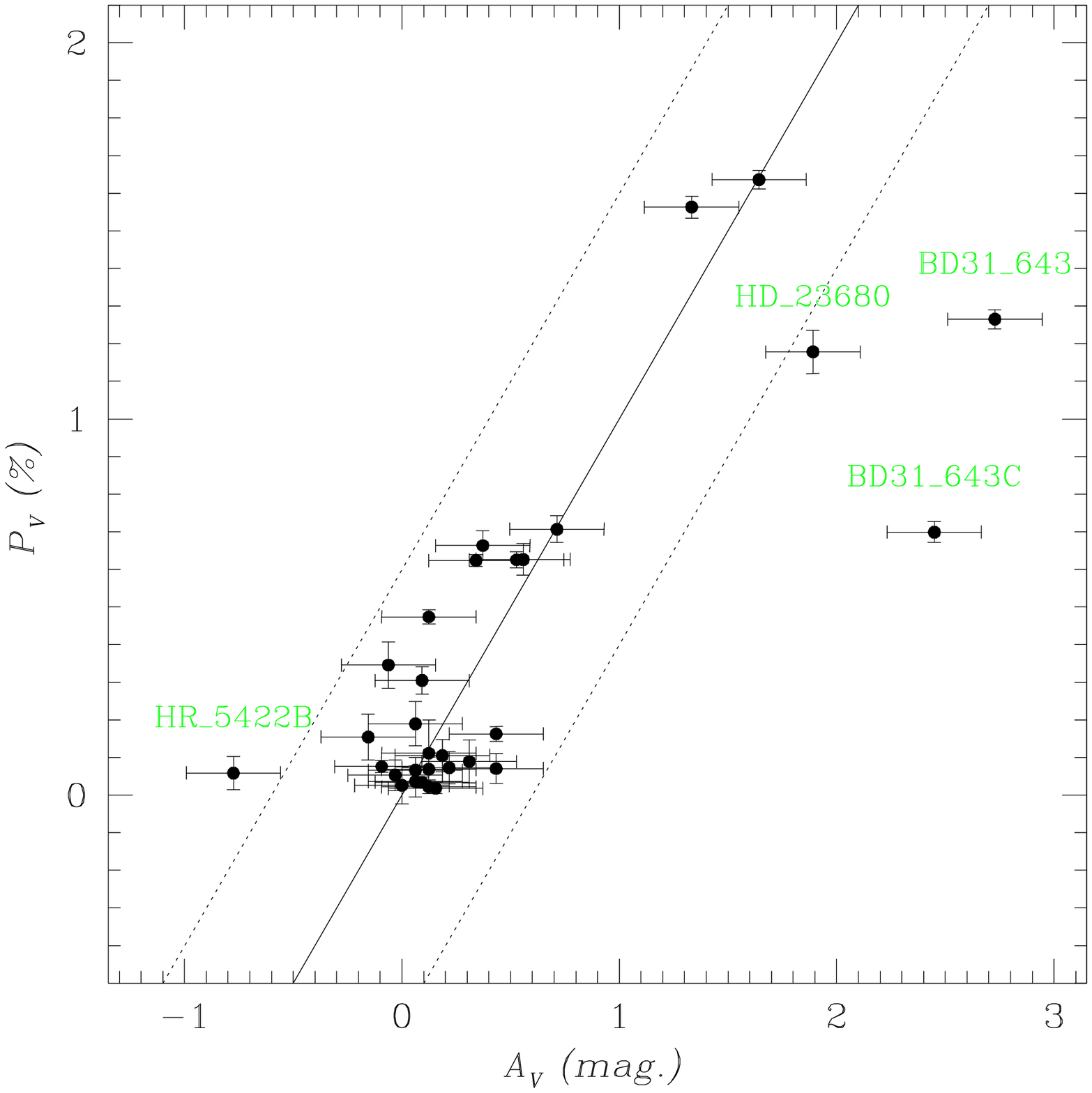}}
\caption{Observed polarization in the {\it V} band of the Vega-type stars plotted against their derived $A_V$. The solid line represents the relation $P_V (\%) \sim A_V$, while the parallel dotted lines represent 3$\sigma$ error bars on the derived $A_V$.
\label{vegaplot}}
\end{figure}

It is a rather well established finding that the interstellar
polarization (ISP), which arises from the presence of aligned dust
grains in the interstellar medium is correlated with the extinction
due to the same dust grains. For example Serkowski et al (1975) showed
that the maximum polarization (in \%) is always smaller than 9 $\times
E(B - V)$ (in magnitudes). In fact, there is a relatively strong
correlation between the observed polarization and extinction in such a
way that $P_V (\%) \sim A_V$.  For example, the sample of 361 field
stars observed by Reiz \& Franco (1998, excluding two objects that
have no derived reddening values) gives an average ($P_V - A_V$) of
0.09 with a scatter of 0.23. The scatter is larger than the estimated
errors on the measured polarizations and derived reddenings by Reiz
\& Franco.

Hence, a useful test would be to look for sources that have
excess polarization compared to that expected from their observed
$A_V$. The polarization could then very well be due to a dusty disk,
which does not contribute to the extinction, as an inclined disk would
let the observer look straight at the star. To this end, we calculated
the reddening to each of our objects, using our {\it (B -- V)}
photometric data, spectral types taken from Mora et al (2001), who
obtained spectral types from spectra taken simultaneously with the
photo-polarimetry in our EXPORT programme, and assuming a normal
reddening law $A_V = 3.1E(B - V)$. $E(B - V)$ was calculated using
intrinsic {\it (B -- V)} colours listed in Schmidt-Kaler (1982) for
the spectral types. For those objects that were not spectrally typed
by EXPORT, or for which no photometry was available, the data provided
by SIMBAD were used. The results are listed in Table~\ref{vegas},
along with weighted mean values for the polarization and polarization
angles in the {\it V} band. The estimated error in $A_V$ is about 0.2
mag, a result from the propagation of errors in the photometry
(assumed to be 0.05 magnitudes for each band for each star) and the
uncertainty in the spectral type (mostly two subclasses - Mora et al)
which results in an average uncertainty in {\it (B -- V)$_0$} of about
0.08 mag. The uncertainty in the luminosity class was not taken into
account, only for G and K stars this would be larger.

The observed polarization as function of the derived $A_V$ is shown in
Fig.~\ref{vegaplot}, where the relation $P_V (\%) \sim A_V$ is shown
as a solid line, and the dotted lines represent a 3$\sigma$ error bar
in this relation (0.6 mag -see above). Most of the stars in the sample
indeed fall within these boundaries, in fact, if we exclude the 
objects that fall either above or below these lines, we find an
average $P_V - A_V$ of 0.05 with an root-mean-square scatter of
0.19, very close to what is found for Reiz \& Franco's 1998 data.

The `excess' polarization of HR 5422B ($P_V$ = 0.06 $\pm$ 0.04 \%)
stands out because of its very negative $A_V$ value. Inspection of the
photometry shows that  the data quality is good. 
We note that the spectral type should be changed from K0 to G0 in
order to arrive at a different intrinsic {\it B -- V} colour or change
the observed {\it B -- V} to 0.85 to alleviate this problem.  The
spectrum of the object shows hints of veiling (Mora, private
communication) which may conspire with a possible UV-excess to result
in too blue observed colours for its spectral type.

We also find three objects which show less polarization than expected,
HD 23680, BD +31\degree643, and BD +31\degree643C. This would not be
exceptional, as the observed relation between polarization and
extinction should be regarded as an upper limit, but it could reveal
the presence of a second polarization mechanism, destructively adding
to the total polarization due to interstellar dust.

\subsection{Deviations from Serkowski law}

Although it can now be argued that all objects only suffer ISP as they
are located on the expected relation between ISP and interstellar
extinction, our multiwavelength data can offer ways to assess whether
intrinsic polarization does contribute a fraction to the observed
polarization.

We therefore continued with the sub-sample of Vega-like stars that are
polarized, and removed objects with observed polarization less than
the 3$\sigma$ level at more than one band. These stars are indicated
with 3$\sigma$ upper limits in Table~\ref{vegas}. HR 419 
was only observed to be polarized in the {\it V} band, and
is also removed for the further study.

\subsubsection{The polarization angle as function of wavelength}

A first indication is to investigate whether the polarization angle,
$\theta$, changes over the different pass bands. Normal interstellar
polarization always displays the same position angle, as the grains
responsible are aligned in the same manner. Any deviations from this
may reveal the presence of an intrinsic component.  We calculated, in
much the same way as we detected variability in the polarization the
r.m.s. scatter of the polarization angle at each wavelength around the
mean value. This was then compared  with the experimental
error, set at minimally 0.5\degree \ in line with the systematic
errors expected. Two objects have a variability, computed using Eq.~1,
in their polarization angles much larger than three. These are BD
+31\degree643
and BD+31\degree643C.

\begin{figure*}
\mbox{\epsfxsize0.245\textwidth\epsfbox{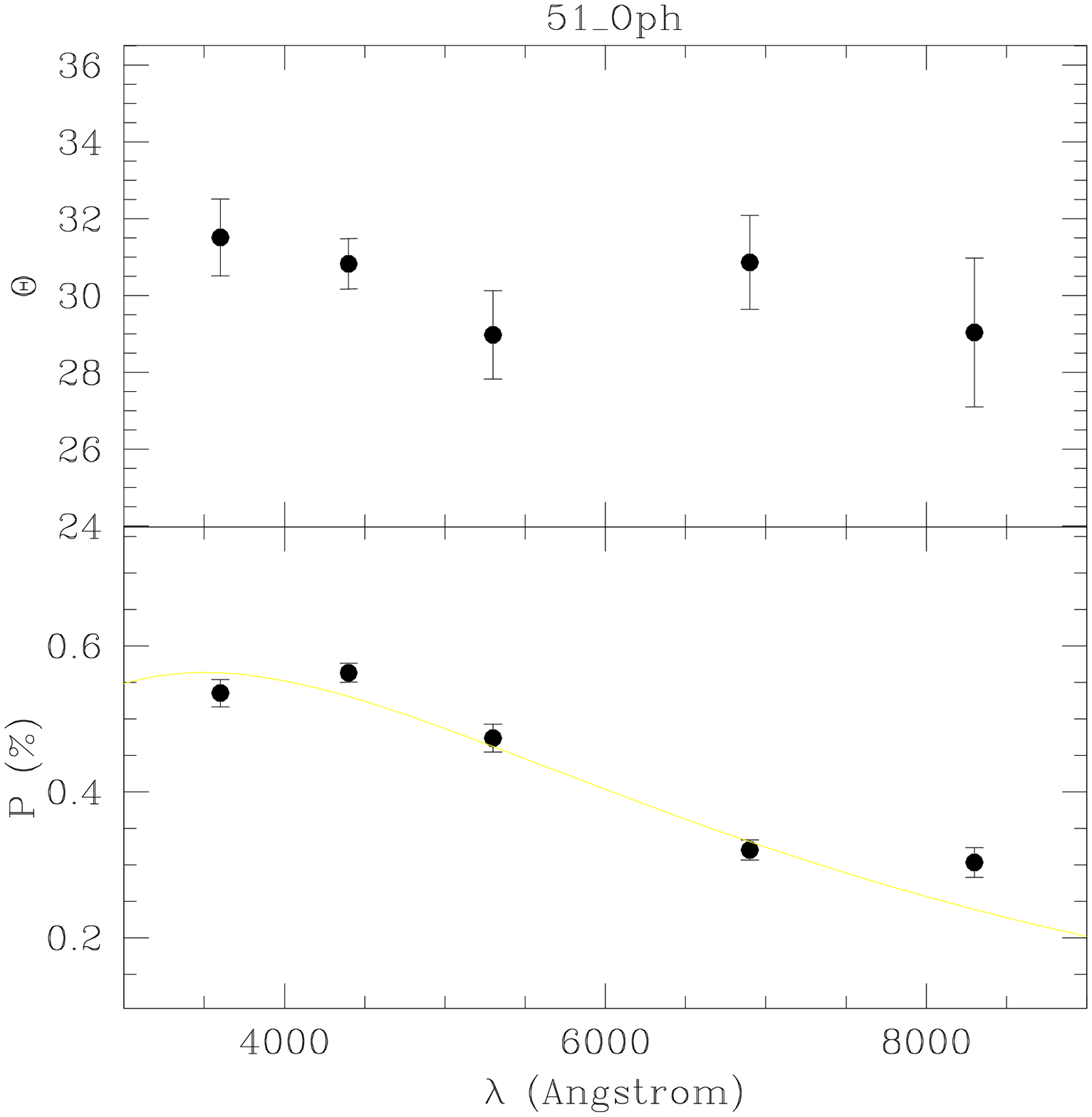}}
\mbox{\epsfxsize0.245\textwidth\epsfbox{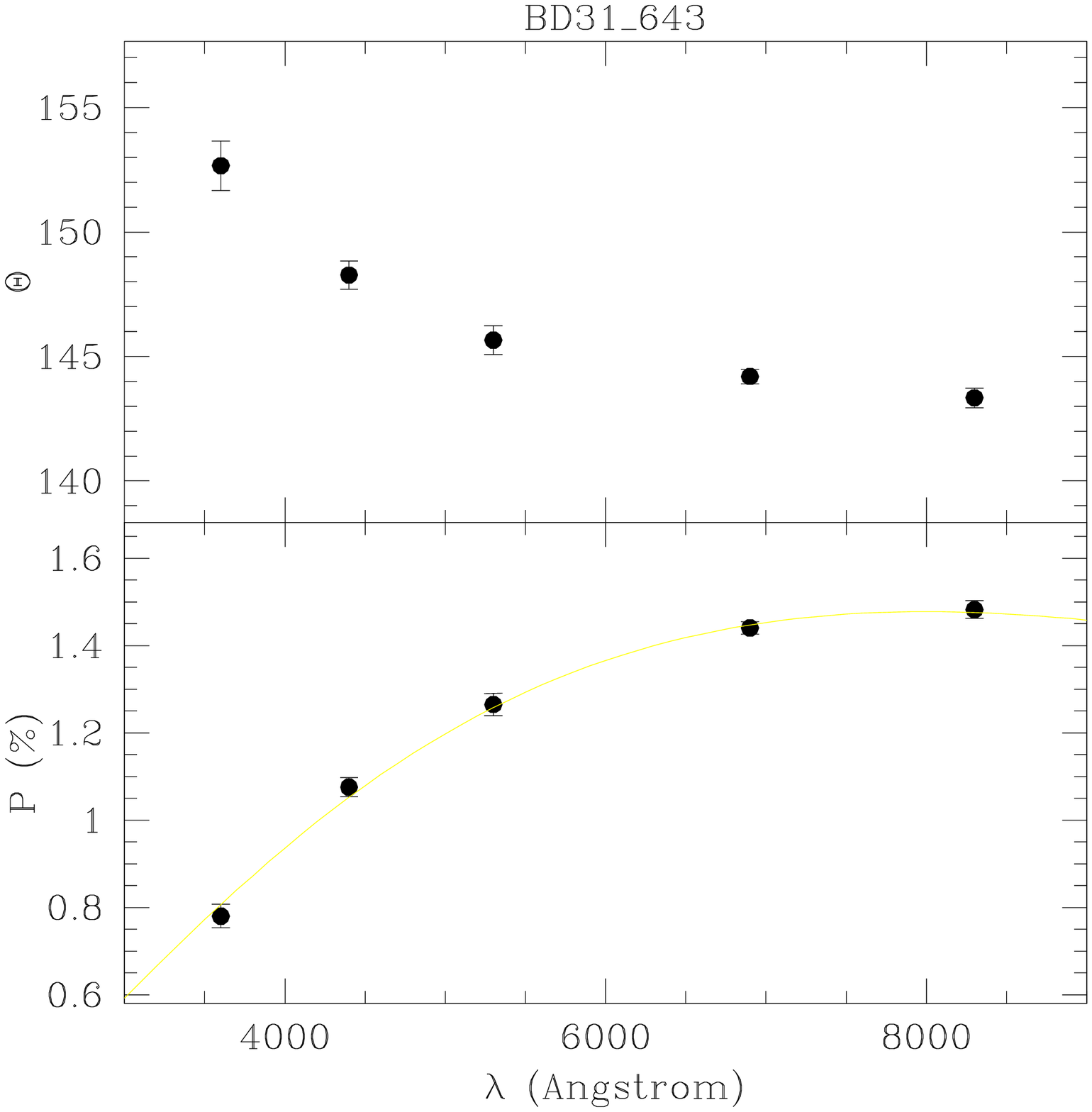}}
\mbox{\epsfxsize0.245\textwidth\epsfbox{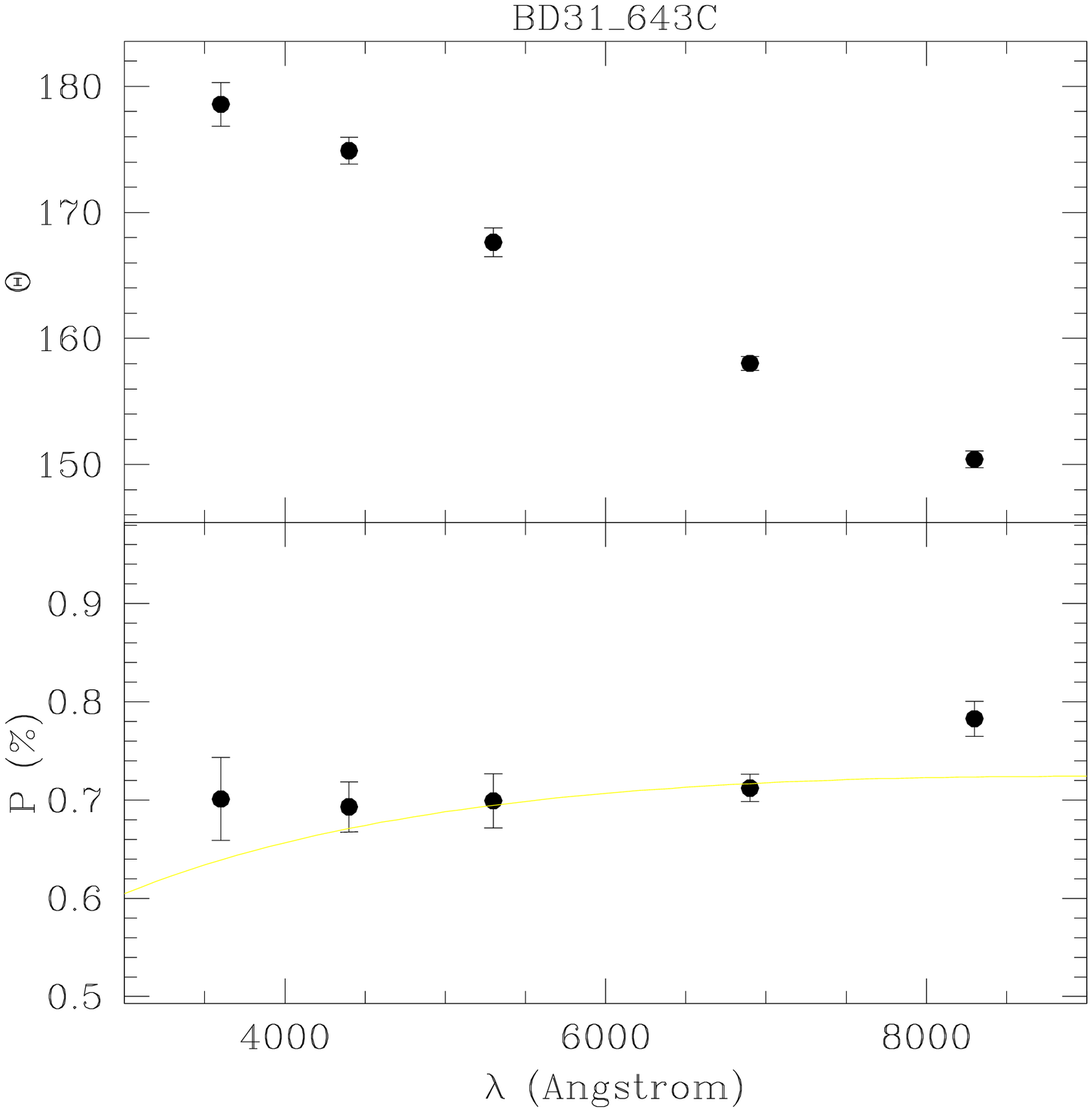}}
\mbox{\epsfxsize0.245\textwidth\epsfbox{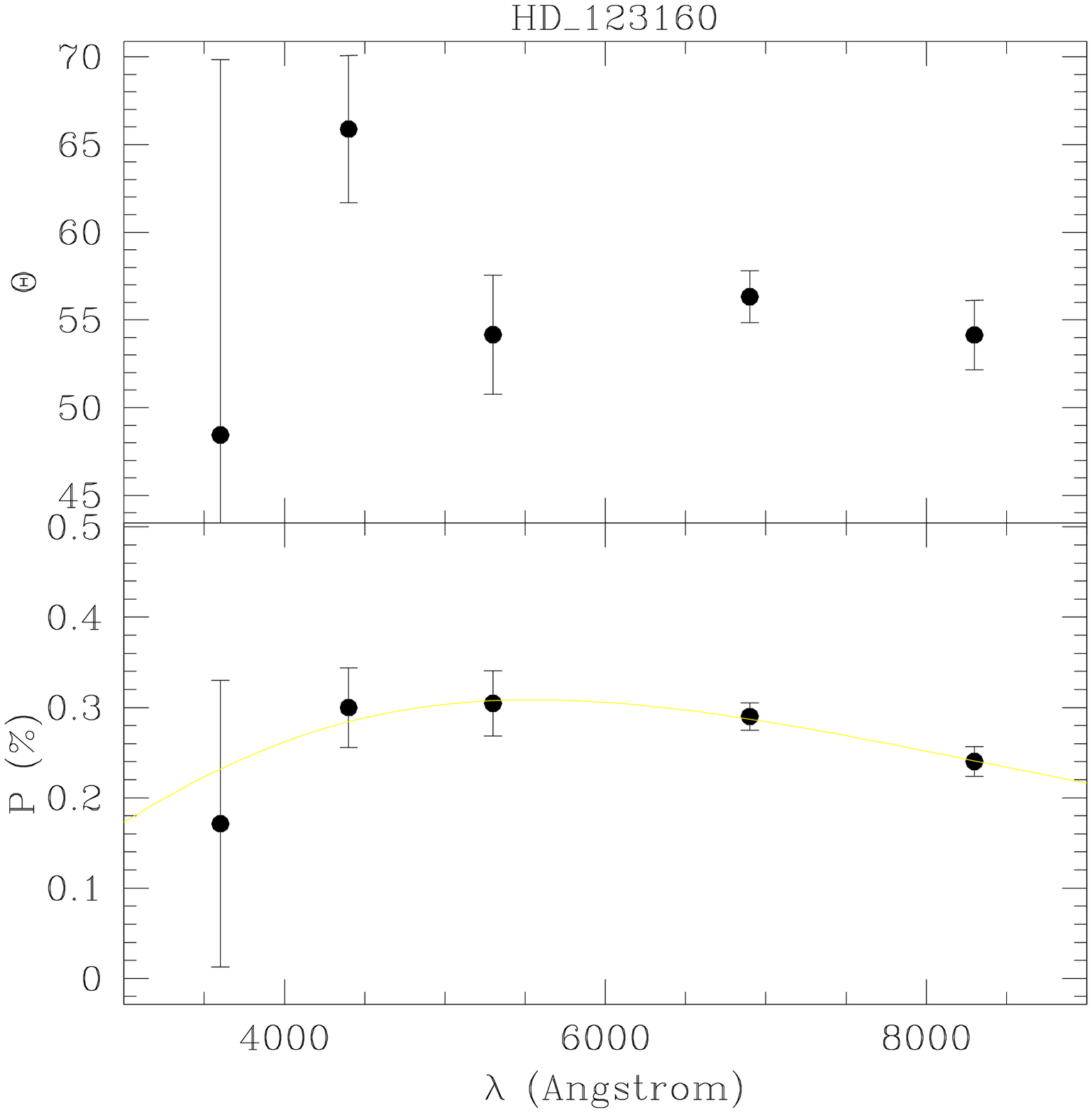}}

\mbox{\epsfxsize0.245\textwidth\epsfbox{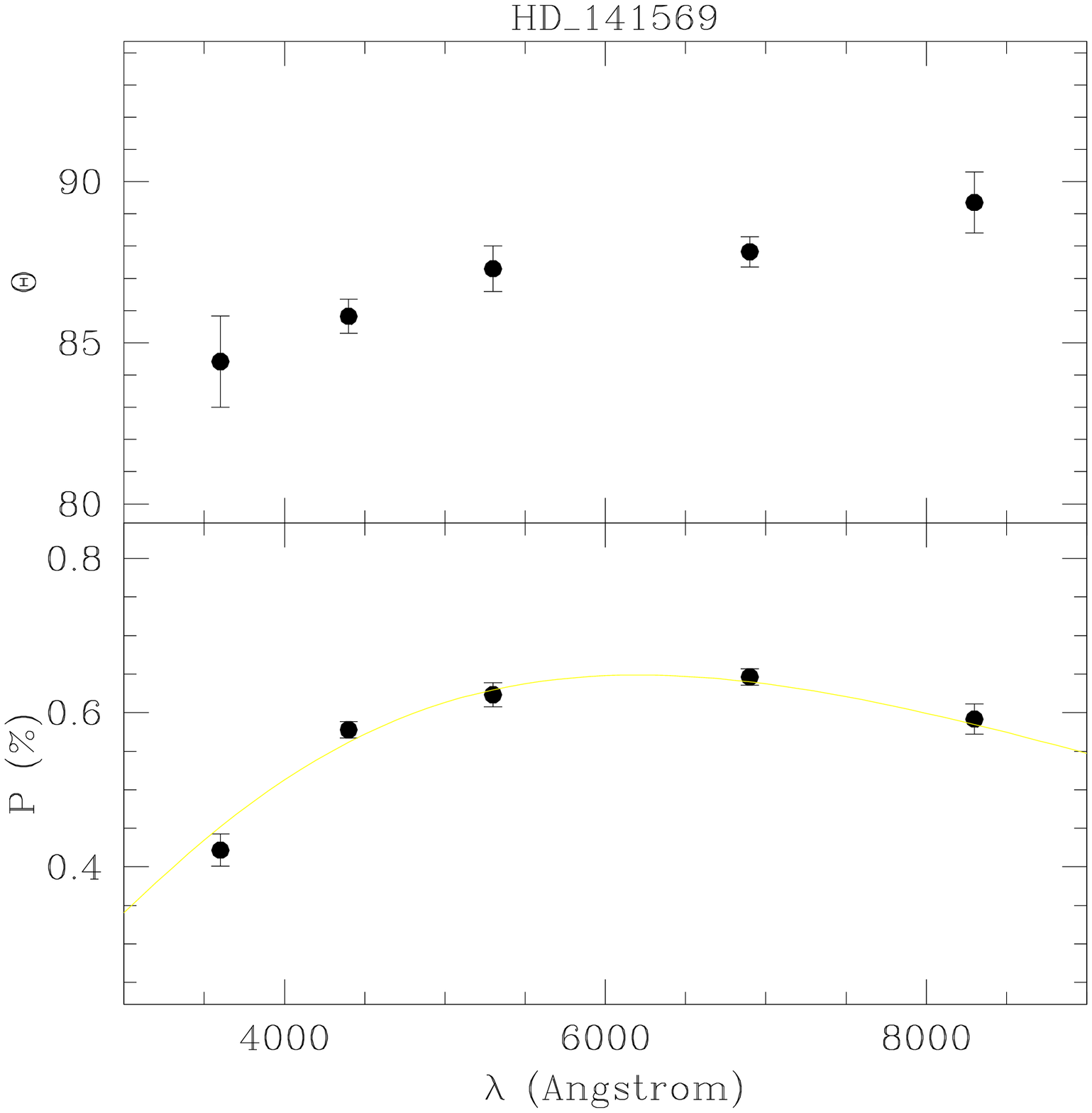}}
\mbox{\epsfxsize0.245\textwidth\epsfbox{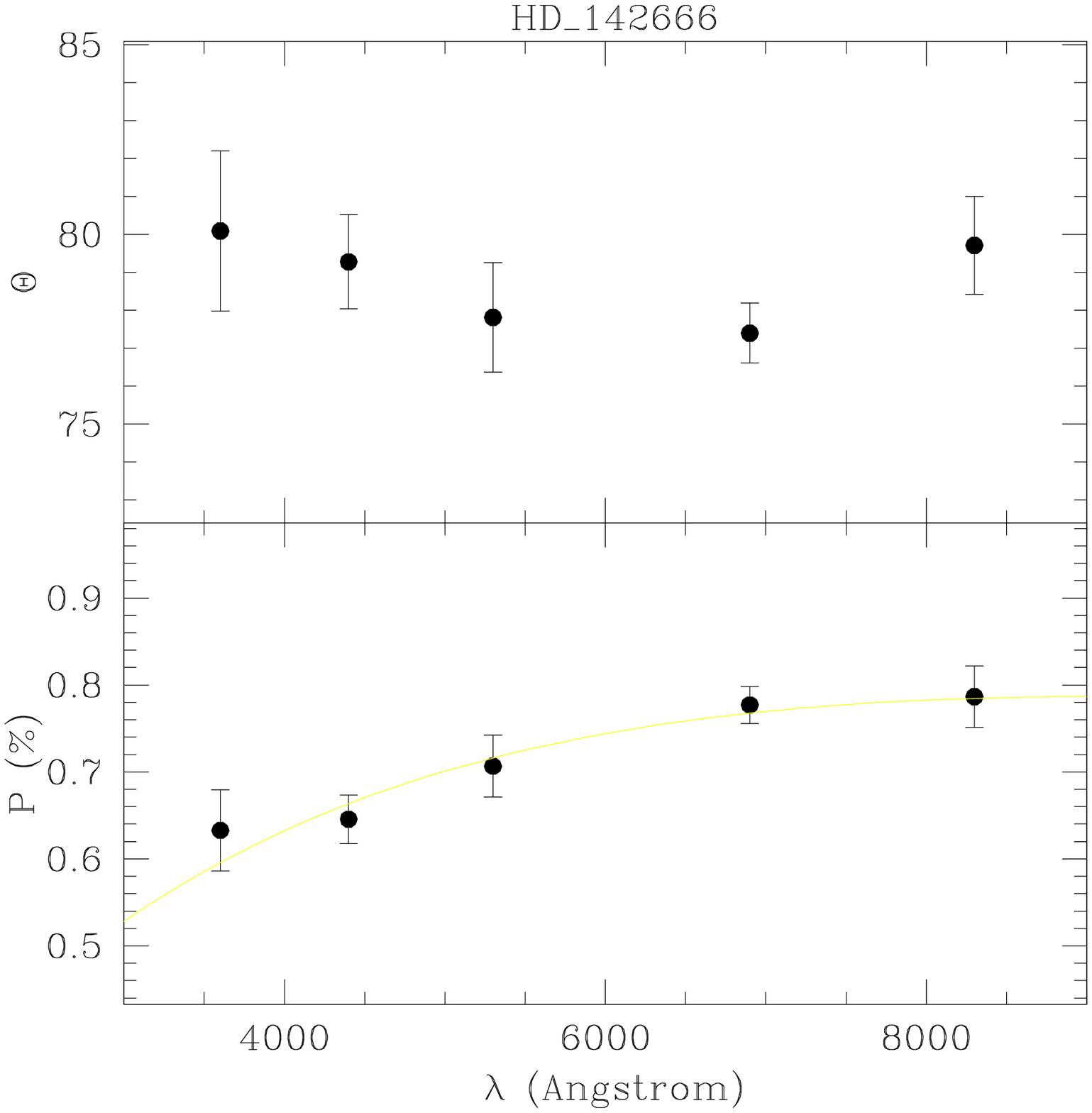}}
\mbox{\epsfxsize0.245\textwidth\epsfbox{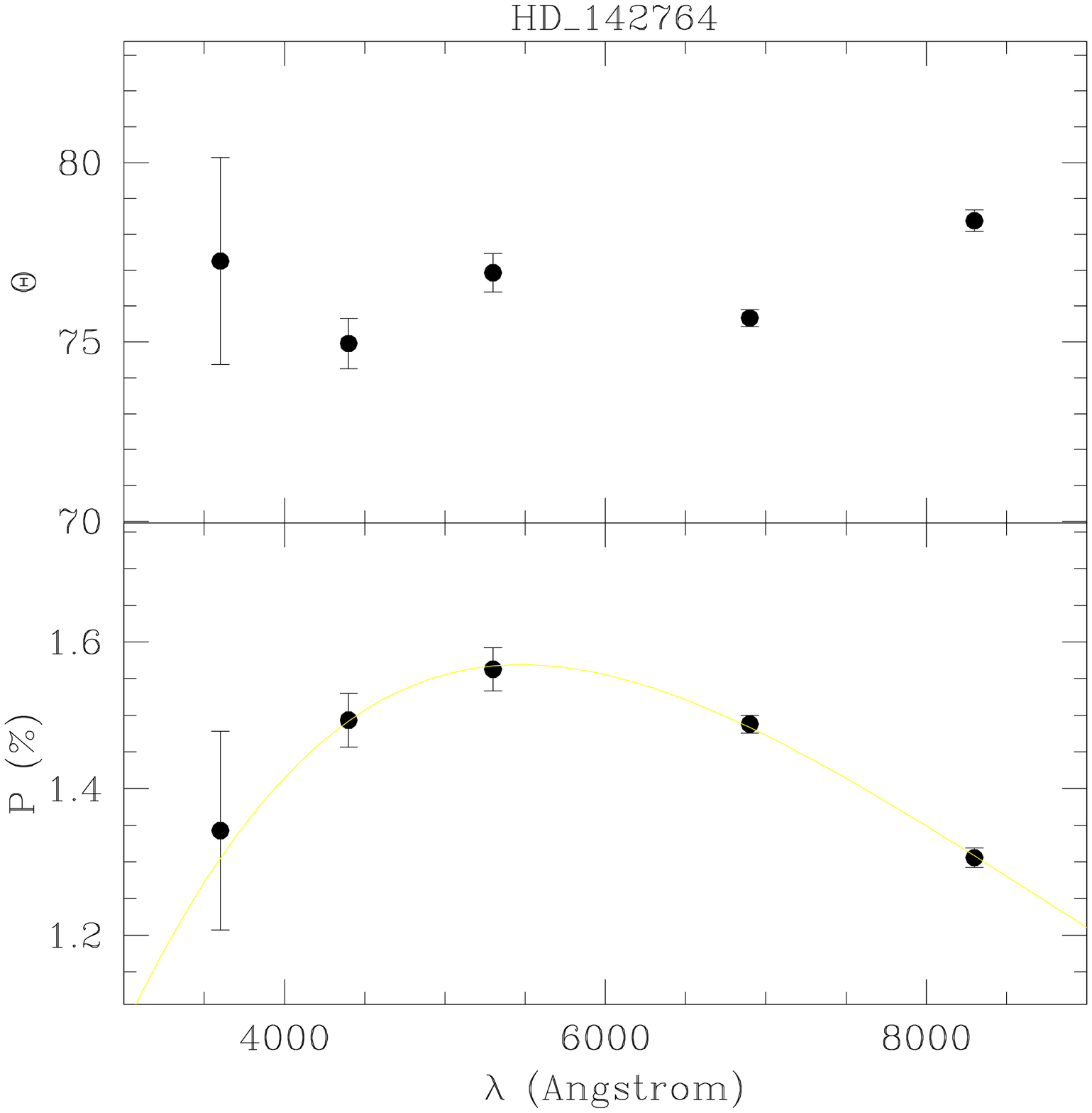}}
\mbox{\epsfxsize0.245\textwidth\epsfbox{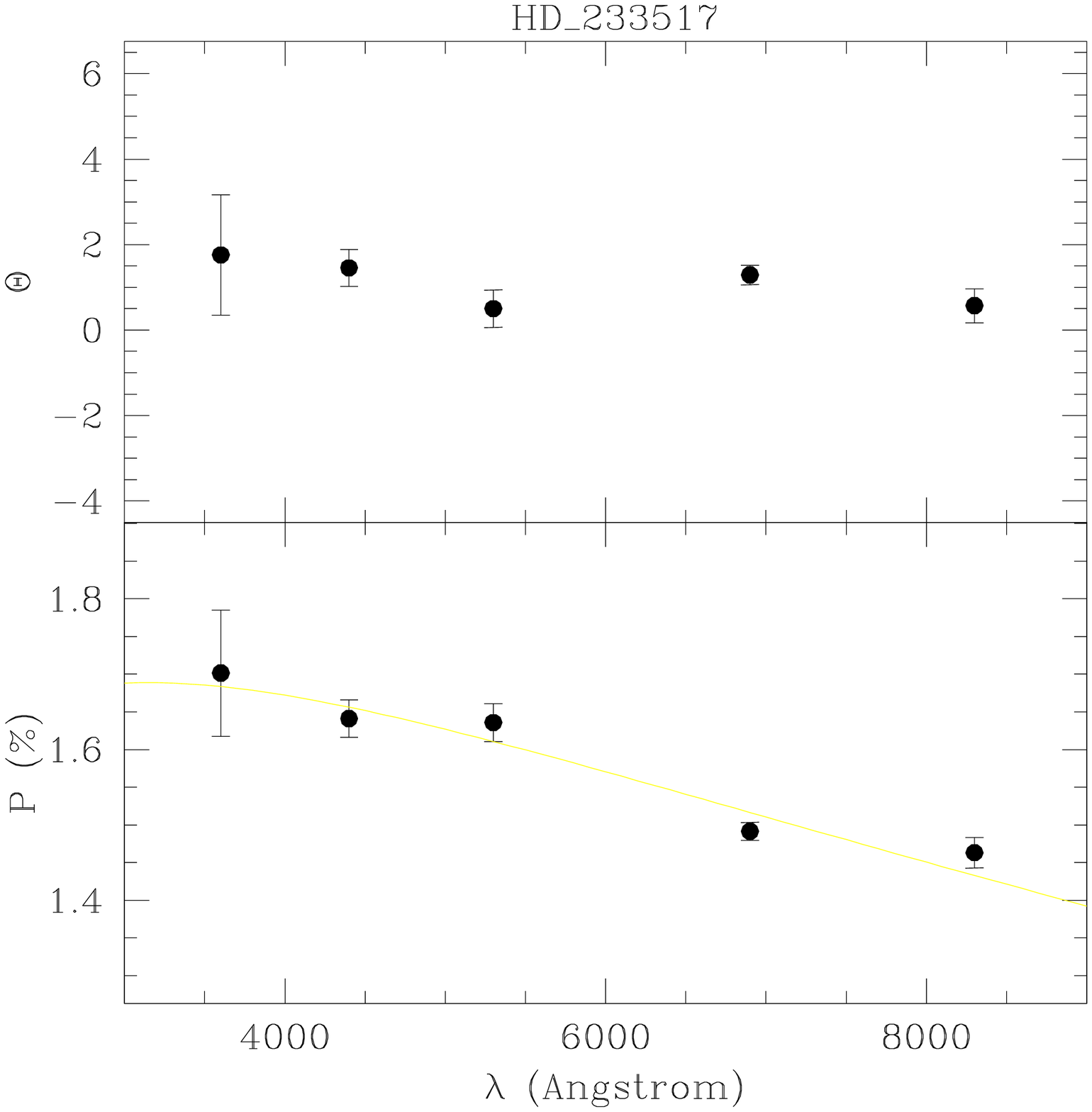}}

\mbox{\epsfxsize0.245\textwidth\epsfbox{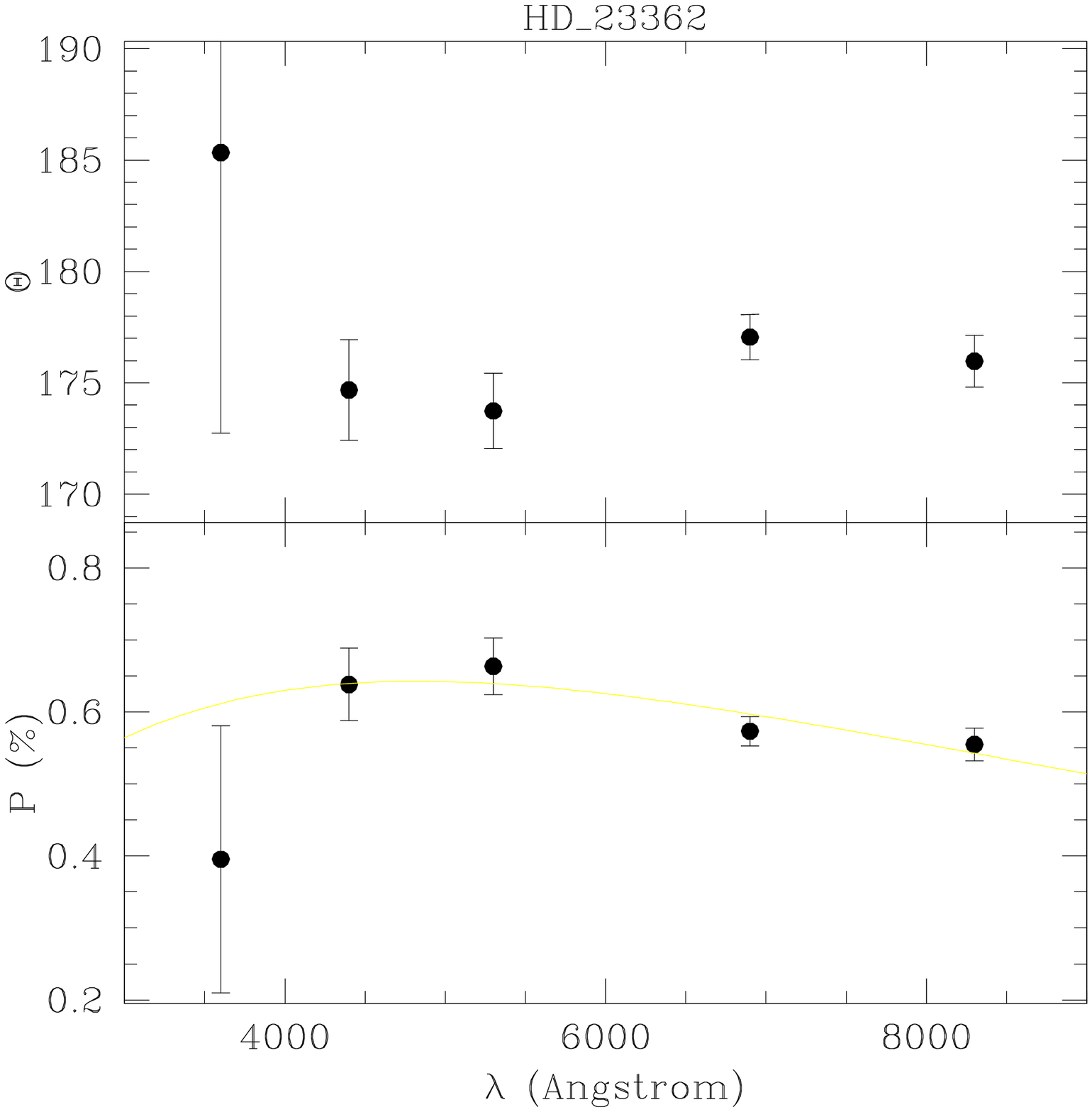}}
\mbox{\epsfxsize0.245\textwidth\epsfbox{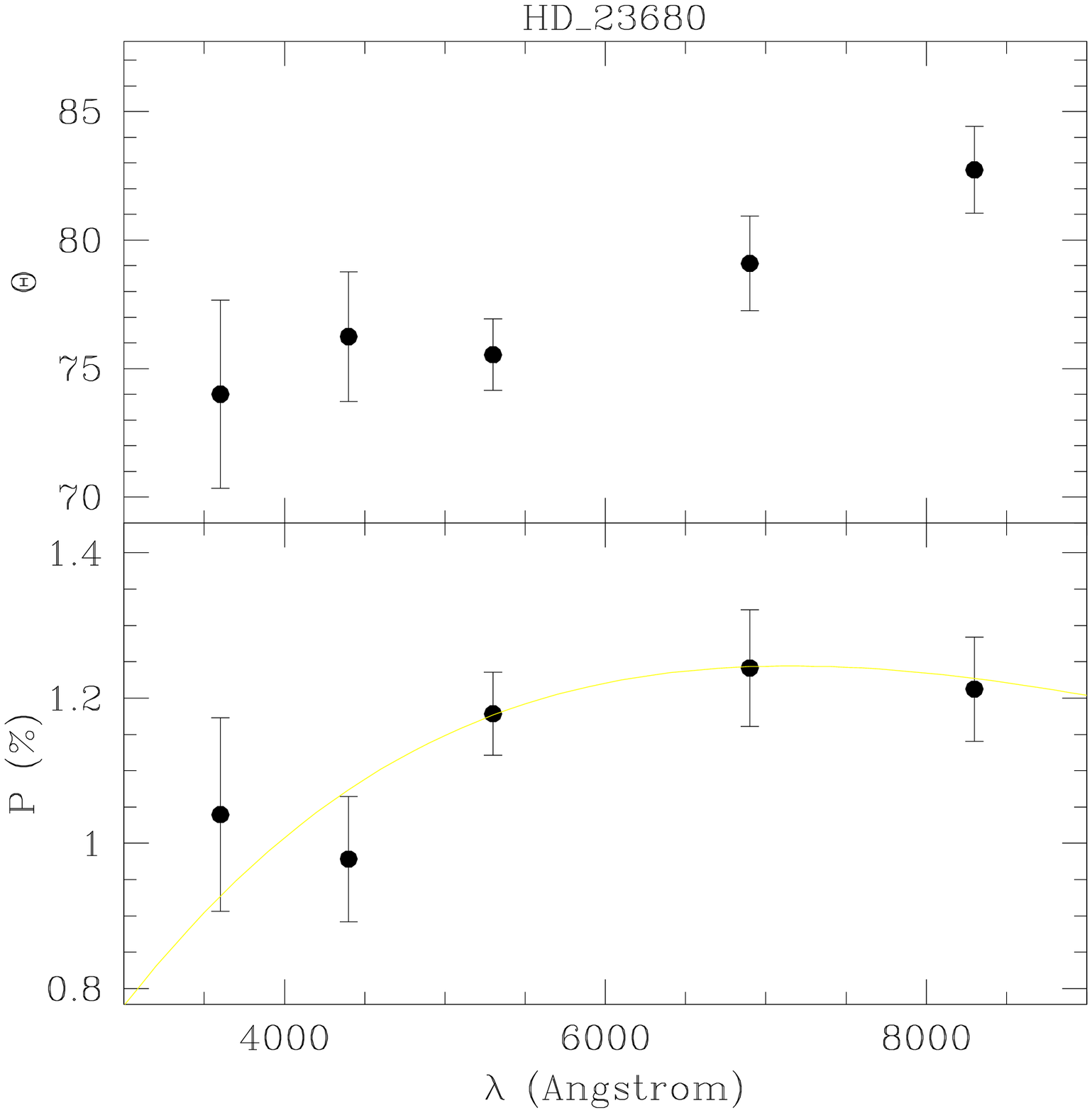}}
\mbox{\epsfxsize0.245\textwidth\epsfbox{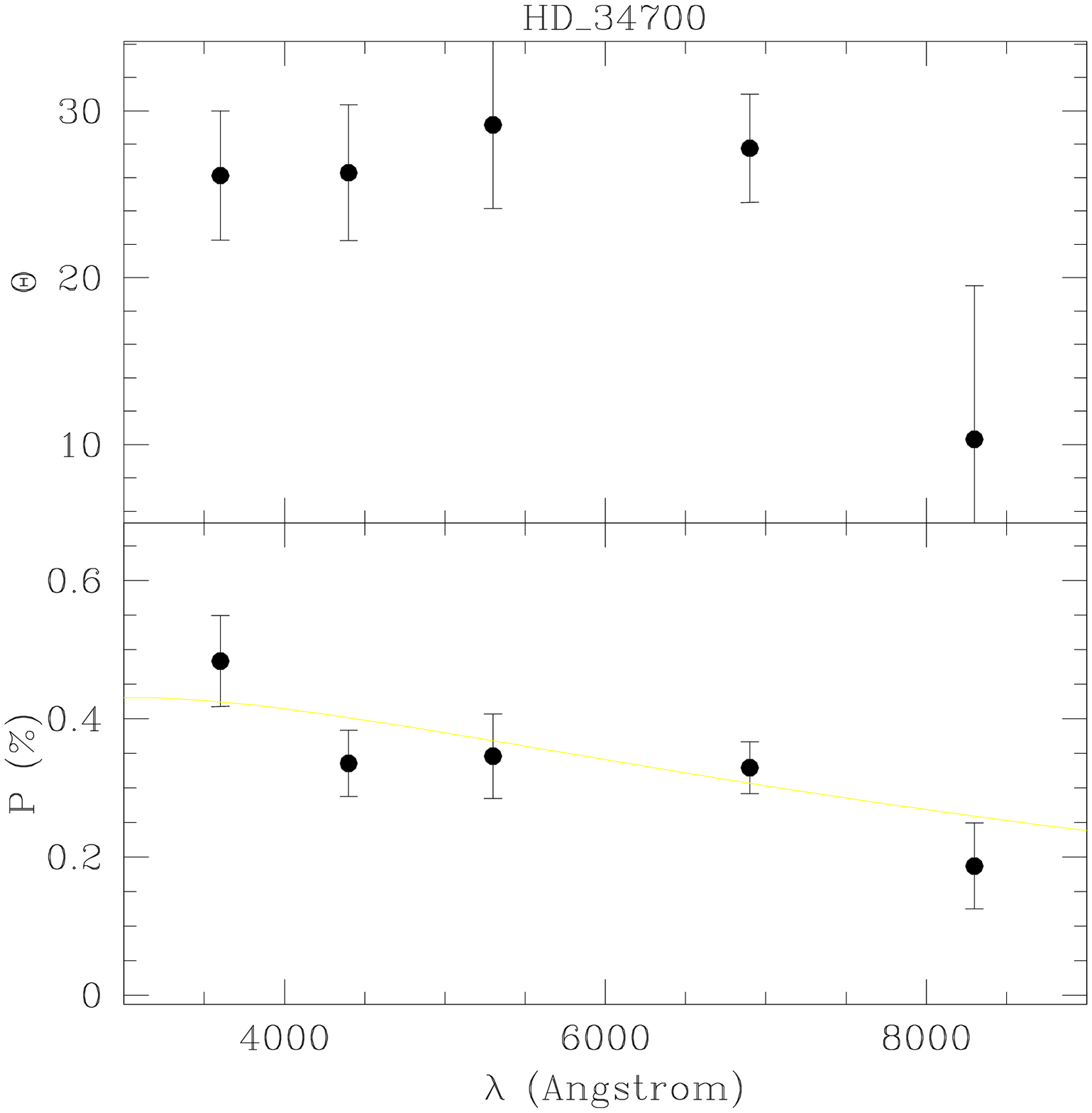}}
\mbox{\epsfxsize0.245\textwidth\epsfbox{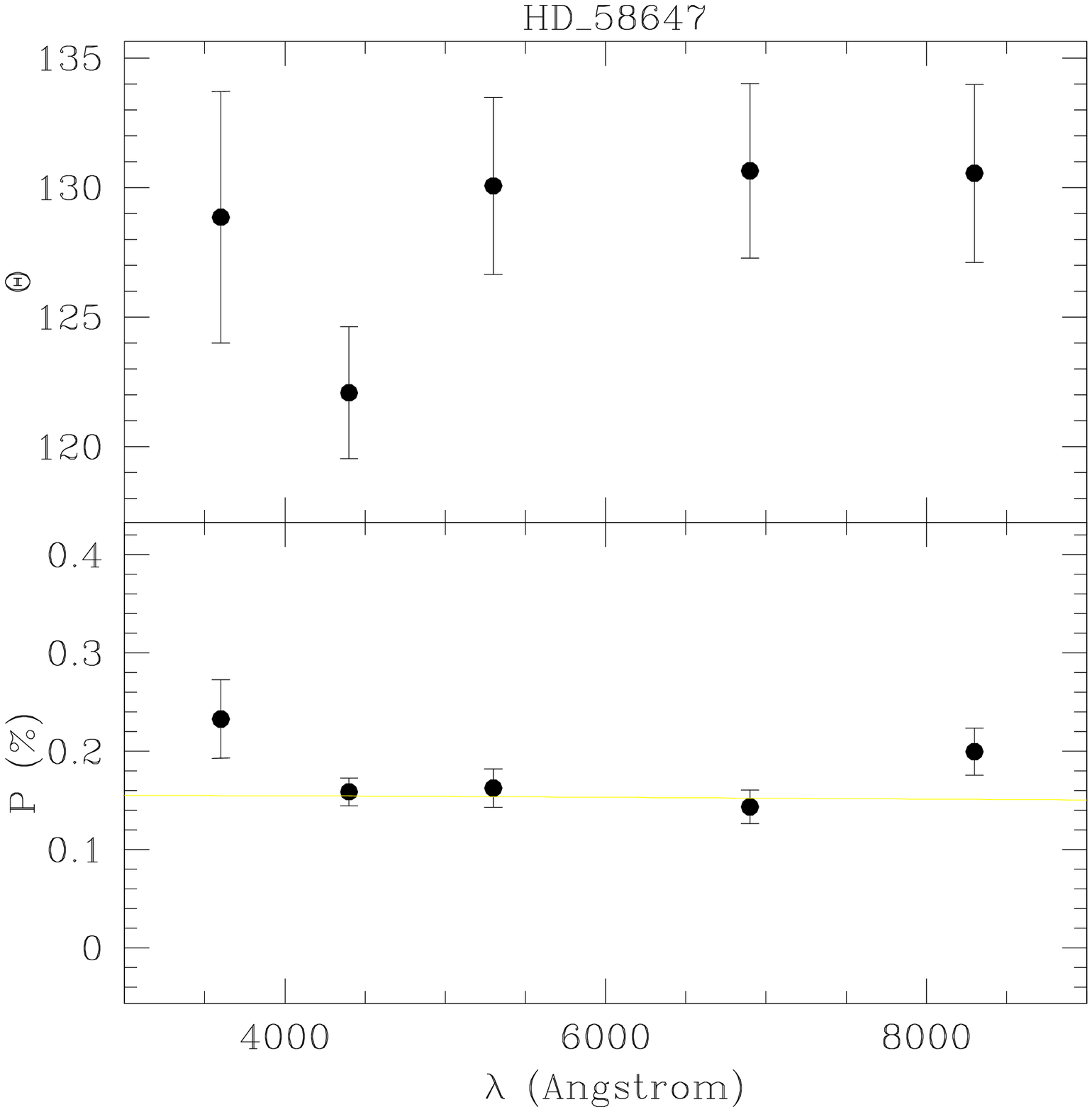}}

\mbox{\epsfxsize0.245\textwidth\epsfbox{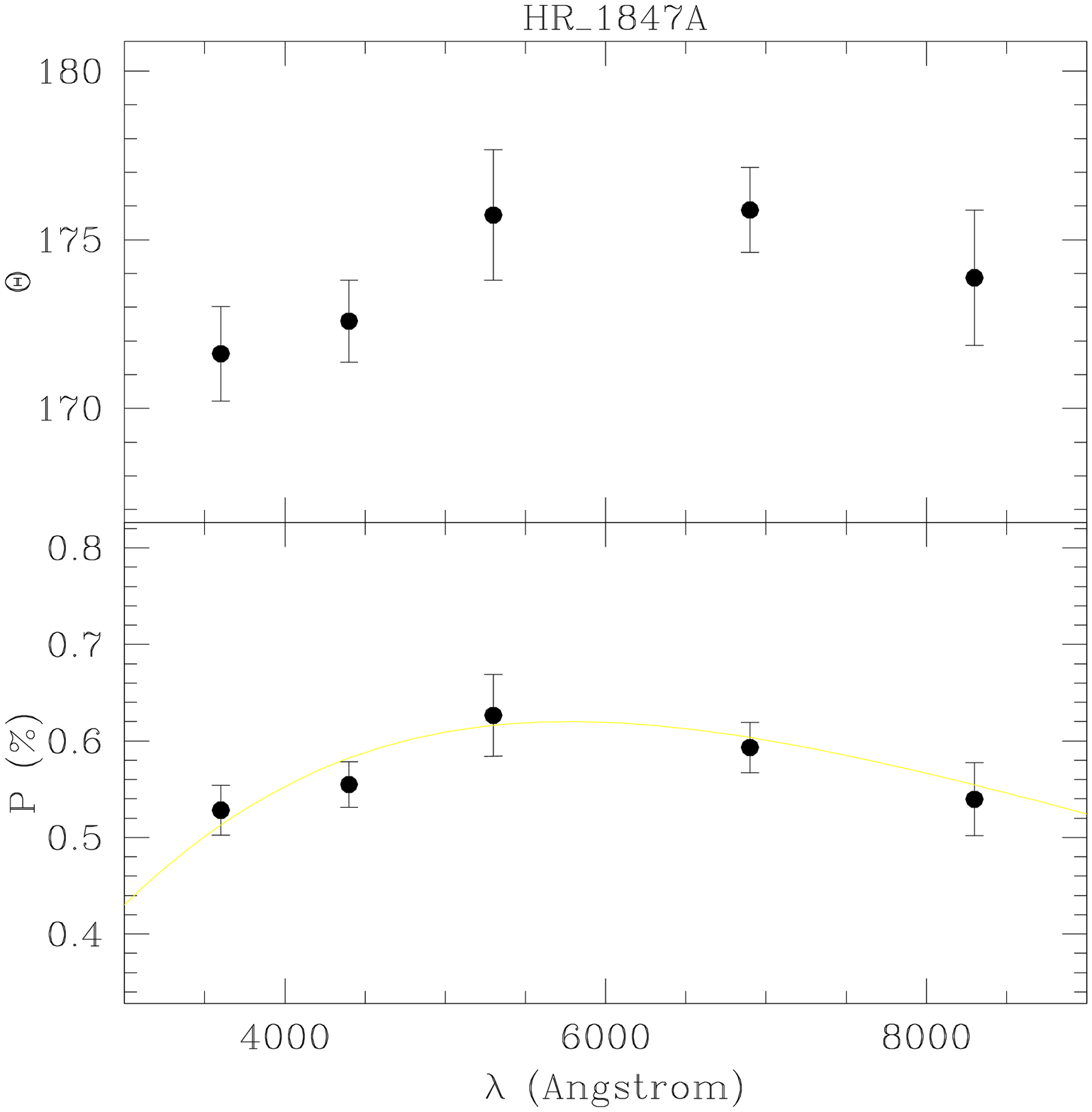}}
\mbox{\epsfxsize0.245\textwidth\epsfbox{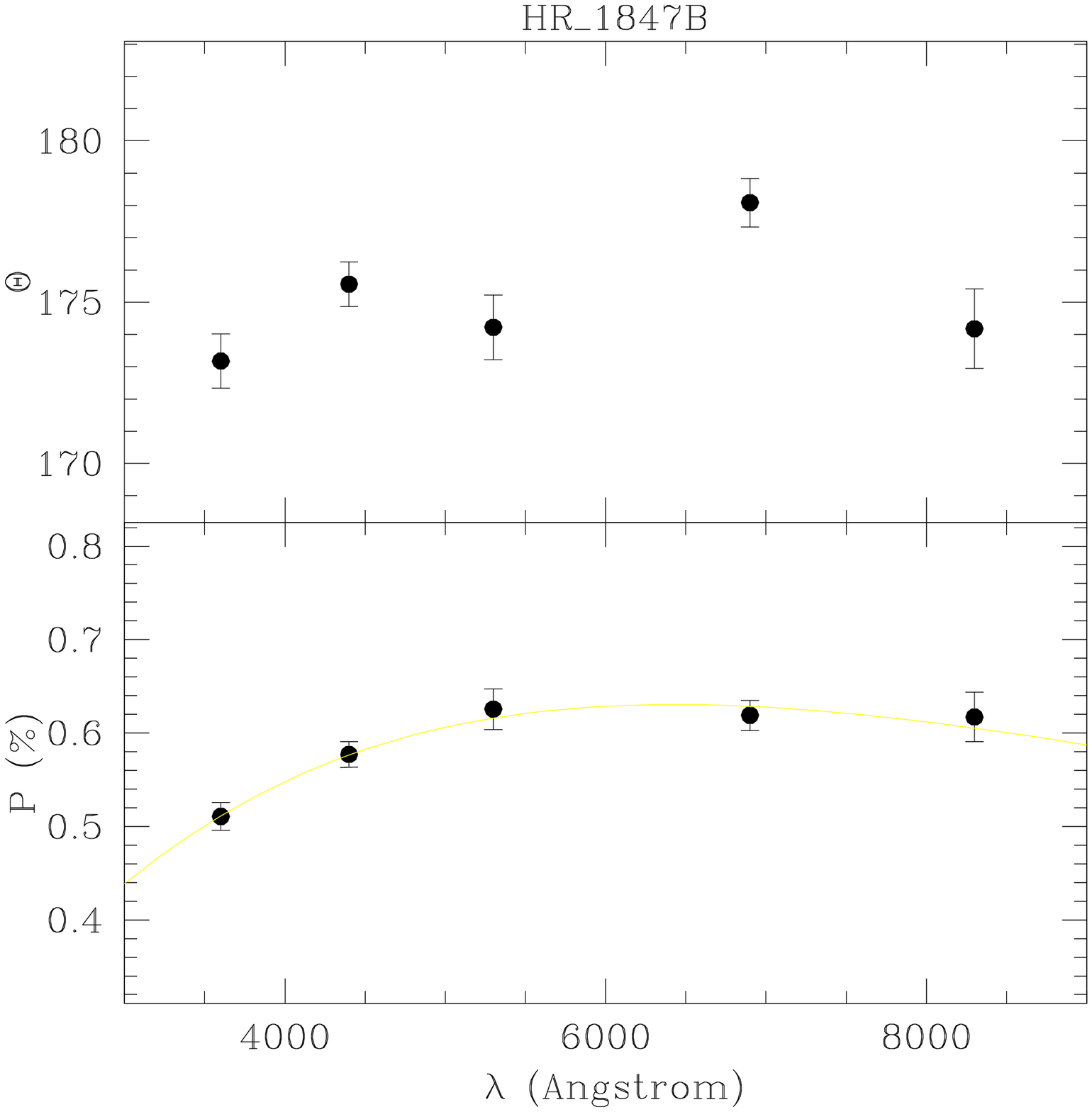}}
\mbox{\epsfxsize0.245\textwidth\epsfbox{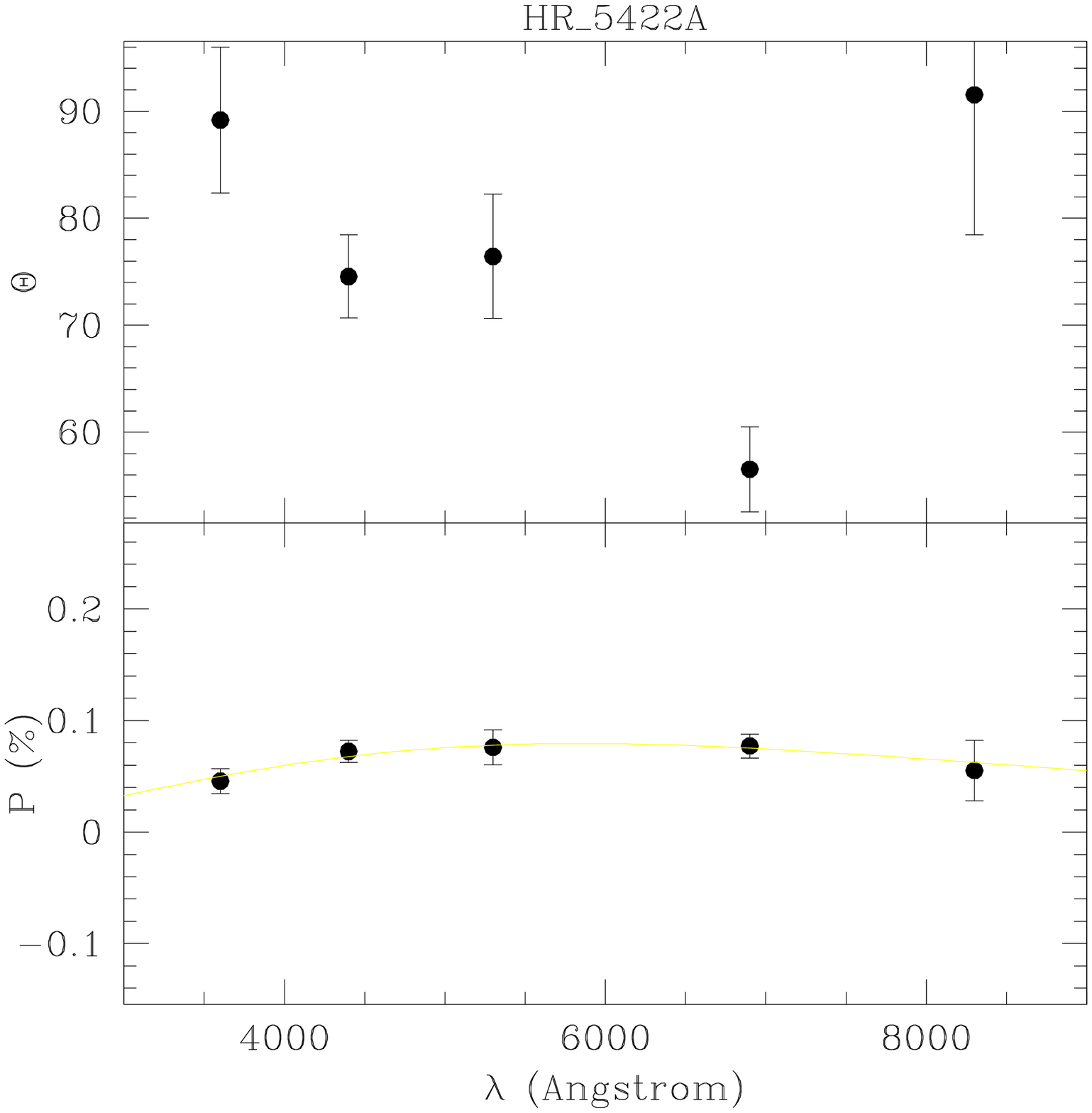}}
\caption{
Weighted mean values of the Polarization and PA as function of wavelength for the Vega-type stars in our sample. The solid lines denote the best fit  Serkowski laws through the points (see text).
\label{vegserk}}
\end{figure*}

\subsubsection{Wavelength dependence of polarization}

A further clue as to whether the observed polarization may be due to
an (additional) intrinsic component is to check whether the data can
be realistically fitted with a Serkowski law. As is well known, the
wavelength dependence of ISP can be empirically expressed as

\begin{equation}
p_{\lambda} = p_{\rm max} {\rm exp}[-K{\rm ln}^2(\lambda_{\rm max}/\lambda)]
\end{equation}

\noindent
with $p_{\rm max}$ the maximum polarization occurring at wavelength
$\lambda_{\rm max}$, and $K$ describes the width of the Serkowski
relation, originally set by default at 1.15, it is now commonly used
as a variable in parameterizing the broad-band polarization behaviour
of stars (see e.g. Whittet et al 1992 and 2001 for a thorough
investigation on the subject). Normal interstellar material has
$\lambda_{\rm max}$ in the range between 4000 and 8000 \ang \, and $K$
ranges between 0.5 and 1.5. Roughly speaking, $\lambda_{\rm max}$ traces
the dominant size of the polarizing dust grains, while $K$ traces the
width of the grain-size distribution. Deviations from this observed
law may also indicate the presence of an intrinsic component, and this
is what we investigate here.  We therefore fitted the data of
the polarized Vega-type objects with this Serkowski law, and searched
the parameter space for the best fit, we let $\lambda_{\rm max}$ run from
3000 to 9000\ang \ and $K$ from 0 to 2. The results of these fits are
plotted in Fig.~\ref{vegserk} and also presented in Table~\ref{vegas}.

Based on the reduced $\chi^2$ computed, we immediately see that
despite having generous limits for searching within $\lambda_{\rm
max}$ and $K$ parameter space, bad fits were obtained for 51 Oph,
BD+31\degree643C, and to a lesser extent HD 58647 and HD 233517 which
have reduced $\chi^2$ slightly larger than 2. It should also be noted
that the best fits of all 4 objects were reached on the borders of the
($\lambda_{\rm max}, K$) values explored, implying that the wavelength
dependence of their polarization can not be due to normal interstellar
material. The latter is also found for other objects, e.g. HD 142666
($\lambda_{\rm max} = 9000$\ang, $K$=0.32), however the small formal
$\chi^2$ (due to the larger observational error bars), provides leeway
for such objects to be satisfactorily represented by a more common
combination.  We note that the presence of two differently polarizing
interstellar dust clouds might also result in a rotation or a
different appearance of the polarization, however, the chances of this
are probably slim, as the Serkowski law was derived from observations
of such stars in the first place.

\subsection{Discussion on Vega-type stars}

Fifteen objects in our sample of 31 Vega-like systems exhibit, within our
error bars, polarization. From simple tests on their
polarization properties, we find that for most objects the observed
polarization can be explained as normal interstellar polarization. We
find evidence that five of these may have an intrinsic component in their
polarization. These stars are discussed individually below:

{\it BD+31\degree643:} BD+31\degree643  shows a trend in
polarization angle which decreases by about 10\degree \ from the {\it U}
band to {\it I}. This was noted already by Andersson \& Wannier
(1997), who interpret this as an additional contribution of
Rayleigh-scattered light from the circumstellar disk to the ISP.

{\it BD+31\degree643C:} This object is located less than 0.4 arcmin
from BD+31\degree643, and is also located within the IC 348
nebula. Little is known about this  bright ({\it V $\sim 10$}) object. Being
picked up by all three of our diagnostic tests, our data show
unambiguously that the polarization of this object can not be due to
simple interstellar polarization alone. The different polarization
behaviour of this star, being almost flat as opposed to the
polarization of BD+31\degree643 that peaks at long wavelengths, leaves
us room to speculate. Assuming the objects were formed in the same
cloud collapse, the later spectral type of
BD+31\degree643C indicates a lower mass, and hence an earlier phase of
formation. Contrary to BD+31\degree643, the polarization of
BD+31\degree643C does not fall off  towards shorter
wavelengths, implying that the smallest dust grains have not yet been
removed during the star+disk evolution.

{\it 51 Oph:} It is especially the increase in polarization towards
smaller wavelengths that indicates this object to exhibit intrinsic
polarization. Taken at face value, the higher polarization at shorter
wavelengths indicates the presence of very small grains, consistent to
the inferences made by van den Ancker et al (2001) from the spectral
energy distribution and features in the ISO-SWS spectrum. The
polarization data clearly imply that the dust responsible for the
infrared emission is located in the scattering disk.

{\it HD 58647:} This object stood out in the Serkowski fitting
technique because of the apparent `excess' polarization in the {\it U}
and {\it I } bands respectively. This behaviour was found on all three
occasions we observed the object. As is the case for 51 Oph, HD
58647 is an H$\alpha$ emitting object, and the possibility that
electron-scattering plays a role can not be excluded. In this case the
deviation from the Serkowski law would not imply a dusty disk-like
structure but a disk-like structure in the ionized gas surrounding the
object.

{\it HD 233517} also has its polarization peak at extremely
short wavelengths. Often included in studies of Vega-type objects,
it now seems that HD 233517 actually is one of the K-giants with
Lithium (Balachandran et al 2000).
The infrared excess of such giants has been interpreted
being due to the evaporation of comets surrounding a Main Sequence
star as it evolves off the Main Sequence (e.g. Plets et al 1997). Our
data indicate that these comet-like bodies may also be distributed in
a disk. We note in passing that recently, not only HD 233517 has been
identified as a Li-rich K-giant with infrared-excess instead of a
Vega-type star, but also HD 23362 (Castilho et al 1998), which was
originally selected from existing lists of Vega-type stars as well.

\section{Discussion and concluding remarks}

We have observed a large sample of low and medium mass main sequence
Vega-type stars and pre-main sequence objects photo-polarimetically
over short (days) and long (months) time-scales. For many of these
objects, these are the first such data obtained in this manner. 

The main global property that we investigated here was the
polarimetric variability of the objects- which confirms the presence
of a flattened circumstellar dusty structure. The presence of
variability was investigated in the data using relatively simple
statistical methods.  The objects classified as UXORs showed
significant variability, both on short and long timescales.
In most cases, the polarimetry is anti-correlated with the photometry:
the objects show larger polarization when they are fainter. This is
generally interpreted in terms of aspherically distributed dust clouds
orbiting the star. These give rise to more extinction in the
line-of-sight, while the scattered light will not be blocked.  This
results in an increase of the relative contribution of the scattered
light to the total observed light, giving rise to larger polarization
(e.g. Grinin et al 1994).

Arguably, the distinction between `UXORs' (T Tauri and Herbig Ae/Be
stars showing large variations in photometry and polarization) and
`normal' T Tauri and Herbig Ae/Be stars becomes less pronounced based
on results like these (see also Herbst \& Shevchenko 1999 for a
discussion on photometric variability). Indeed, this large data-set
indicates that the majority of the observed Herbig Ae/Be stars show
evidence for non-spherical envelopes. Although there may have been a
certain personal bias in the selection of the targets, it should be
noted that most of the observable (northern) known Herbig Ae/Be stars
have been observed in this project.  Considering that the number of
Herbig Ae/Be stars is only about one hundred (Th\'e et al 1994), we
speculate at this point that perhaps {\it all} massive PMS objects
have aspherical structures. This is in line with the random
orientations of any disks in the line of sight. If the disks are
circular, then those disks oriented face-on would not show any
polarization/polarization variability while projected rotational
velocities would, in these cases, be lower.  This is in 
contrast to van den Ancker et al (1998), who did not find a
correlation between photometric variations alone as a function of
rotational velocity of Herbig Ae/Be stars.

Combining our results and the models of Natta \& Whitney may shed some
light on this apparent problem. Natta \& Whitney (2000) argue that if
Herbig Ae stars are surrounded by flaring dusty disks, the UXOR
phenomenon can only be observed for a specific range of inclination
angles.  As outlined above, objects oriented face-on would not exhibit
polarization variability. In addition, objects whose disks are
heavily inclined would not be optically visible, due to the high
optical depth in the disk mid-plane.  Hence, objects with the largest
observed projected rotational velocities are not included in any
optically based sample of Herbig Ae/Be stars, limiting the velocity
range probed by van den Ancker et al. Depending on the (unknown)
opening angle of the dusty disk we may even miss a significant
proportion of existing intermediate mass pre-main sequence stars.  It
may be worthy of note that Natta \& Whitney predict that around half
of the optically visible Herbig Ae stars show the UXOR behaviour, just
as observed here.

The combination of both photometric and polarimetric variability
strongly suggests that all Herbig Ae/Be stars are surrounded by
disks. Of course, we need to analyze the data further, but such a
conclusion, albeit tentative, illustrates the importance of observing
a large sample of PMS stars to be able to reach general conclusions on
the properties of this evasive class of object.

The main result for the more than 30 Vega-type objects is that many
are not polarized to our sensitivity, and those for which polarization
is detected are not variable to within the error bars. Based on simple
arguments, we have flagged 5 objects that may exhibit intrinsic
polarization of which one (BD+31\degree643) was previously known, and
list these as suitable candidates for future research.

\paragraph*{\it Acknowledgments:}

The Nordic Optical Telescope is operated on the island of La Palma
jointly by Denmark, Finland, Iceland, Norway, and Sweden, in the
Spanish Observatorio del Roque de los Muchachos of the Instituto de
Astrof\'\i sica  de Canarias.

\end{document}